\documentclass[lettersize,journal]{IEEEtran}
\usepackage{amsmath,amsfonts}
\usepackage{array}
\usepackage{subfig}
\usepackage{stfloats}
\usepackage{url}
\usepackage{verbatim}
\usepackage{graphicx}
\usepackage{cite}
\usepackage{microtype}
\usepackage{doi}
\usepackage{float}
\usepackage{mathrsfs}
\usepackage{threeparttable}
\usepackage{booktabs}
\usepackage{multicol}
\usepackage{multirow}
\usepackage{tabto}
\usepackage[ruled]{algorithm2e}
\usepackage{algpseudocode}
\usepackage{makecell}
\usepackage{array}
\usepackage{diagbox}
\usepackage{adjustbox}
\usepackage{hyperref}
\usepackage{textcomp}
\usepackage{stackrel,amssymb}
\usepackage{xcolor}

\begin{document}
 \title{Distillation-Driven Diffusion Model for Multi-Scale MRI Super-Resolution: Make 1.5T MRI Great Again}
 
\author{Zhe Wang, Yuhua Ru, Fabian Bauer, Aladine Chetouani, Fang Chen, Liping Zhang, Didier Hans$^*$, Rachid Jennane$^*$, Mohamed Jarraya$^*$, Yung Hsin Chen$^*$
\thanks{Zhe Wang, Yung Hsin Chen and Mohamed Jarraya are with Department of Radiology, Massachusetts General Hospital, Harvard Medical School, Boston, 02114, USA (e-mail: zwang78@mgh.harvard.edu; ychen4@mgh.harvard.edu; mjarraya@mgh.harvard.edu).}
\thanks{Aladine Chetouani is with L2TI Laboratory, University Sorbonne Paris Nord, Villetaneuse, 93430, France (e-mail: aladine.chetouani@univ-paris13.fr).}
\thanks{Yuhua Ru is with Jiangsu Institute of Hematology, The First Affiliated Hospital of Soochow University, Suzhou, 215006, China. (e-mail: ruyuhua@163.com).}
\thanks{Fang Chen is with department of Medical School, Henan University of Chinese Medicine, Zhengzhou, 450046, China. (e-mail: chenfangyxy@hactcm.edu.cn).}
\thanks{Fabian Bauer is with Division of Radiology, German Cancer Research Center, Heidelberg, 69120, Germany (e-mail: fabian.bauer@dkfz-heidelberg.de).}
\thanks{Liping Zhang is with Athinoula A. Martinos Centre for Biomedical Imaging, Massachusetts General Hospital, Harvard Medical School, Boston, 02114, USA. (e-mail: lzhang90@mgh.harvard.edu).}
\thanks{Didier Hans is with Nuclear Medicine Division, Geneva University Hospital, Geneva, 1205, Switzerland. (e-mail: didier.hans@chuv.ch).}
\thanks{Rachid Jennane is with IDP Institute, UMR CNRS 7013, University of Orleans, Orleans, 45067, France (e-mail: rachid.jennane@univ-orleans.fr).}}

\markboth{Journal of \LaTeX\ Class Files,~Vol.~14, No.~8, August~2021}%
{Shell \MakeLowercase{\textit{et al.}}: A Sample Article Using IEEEtran.cls for IEEE Journals}

\maketitle
\begin{abstract}
Magnetic Resonance Imaging (MRI) offers critical insights into microstructural details, however, the spatial resolution of standard 1.5T imaging systems is often limited. In contrast, 7T MRI provides significantly enhanced spatial resolution, enabling finer visualization of anatomical structures. Though this, the high cost and limited availability of 7T MRI hinder its widespread use in clinical settings. To address this challenge, a novel Super-Resolution (SR) model is proposed to generate 7T-like MRI from standard 1.5T MRI scans. Our approach leverages a diffusion-based architecture, incorporating gradient nonlinearity correction and bias field correction data from 7T imaging as guidance. Moreover, to improve deployability, a progressive distillation strategy is introduced. Specifically, the student model refines the 7T SR task with steps, leveraging feature maps from the inference phase of the teacher model as guidance, aiming to allow the student model to achieve progressively 7T SR performance with a smaller, deployable model size.  Experimental results demonstrate that our baseline teacher model achieves state-of-the-art SR performance. The student model, while lightweight, sacrifices minimal performance. Furthermore, the student model is capable of accepting MRI inputs at varying resolutions without the need for retraining, significantly further enhancing deployment flexibility. The clinical relevance of our proposed method is validated using clinical data from Massachusetts General Hospital. Our code is available at \url{https://github.com/ZWang78/SR}.
\end{abstract}


\section{Introduction}
\IEEEPARstart{M}{agnetic} resonance imaging (MRI) is a powerful and versatile medical imaging technique that leverages strong magnetic fields, radiofrequency pulses, and sophisticated computational algorithms to produce detailed images of internal body structures \cite{wolbarst2006evolving}. Its non-invasive nature and superior soft tissue contrast have made it play a pivotal role in diagnosing, monitoring, and managing a wide range of neurological disorders \cite{arabahmadi2022deep}. One of the key strengths of MRI lies in its ability to generate high-resolution images of soft tissues, including the brain, spinal cord, and other critical anatomical regions \cite{mumtaz2022microwave}. This capability has established MRI as indispensable in the detection and characterization of complex neurological conditions such as multiple sclerosis (MS), Alzheimer's disease, Parkinson's disease, brain tumours, and stroke \cite{filippi2004magnetization}. For instance, in MS, MRI is used to visualize demyelinating lesions and monitor disease progression \cite{love2006demyelinating}, while in Alzheimer's, advanced techniques such as volumetric analysis and functional MRI (fMRI) aid in detecting early cortical atrophy and disrupted brain activity \cite{wang2007altered}. Moreover, MRI's role in identifying ischemic and hemorrhagic strokes is critical for timely intervention and improved patient outcomes \cite{zaheer2000magnetic}.

Over the last decade, significant technological advancements have transformed MRI, enhancing its diagnostic power through stronger magnetic fields and higher-resolution imaging capabilities \cite{ai2012historical}. From its early inception in the 1980s, when MRI systems operated at field strengths below 0.5T, the technology has progressed to the widespread clinical adoption of 1.5T systems \cite{blamire2008technology}, and more recently, to the cutting-edge 7T MRI \cite{thomas2008high}. These advancements reflect the remarkable strides in magnet design, gradient performance, and imaging sequences, which have collectively improved image quality, reduced scan times, and expanded the scope of clinical and research applications. The advent of ultra-high-field 7T MRI represents a quantum leap in imaging technology. With its substantially higher spatial resolution and superior signal-to-noise ratio (SNR), 7T MRI enables the visualization of minute anatomical structures and subtle pathological changes that may be overlooked with lower-field systems \cite{pazahr20237}. This capability is critical for improving diagnostic accuracy and prognostic assessments in a variety of neurological conditions \cite{platt20217}. For example, in epilepsy, 7T MRI has demonstrated an unparalleled ability to localize small epileptogenic lesions, facilitating precise surgical planning \cite{de20167t}. Similarly, in multiple sclerosis, it allows for the detection of cortical lesions that are less visible at 1.5T or 3T, providing deeper insights into disease progression and treatment response \cite{kolb20217t}. Despite these advantages, the clinical adoption of 7T MRI remains limited due to several challenges. The high cost of acquisition and installation, coupled with the need for specialized infrastructure such as radiofrequency shielding and cryogenic cooling systems, makes these systems prohibitively expensive for many institutions \cite{kraff20177t}. Moreover, operational expenses, including maintenance and the need for highly trained personnel, further restrict accessibility. As of 2022, fewer than 100 7T MRI scanners were operational worldwide, with most located in well-funded research institutions and specialized centres \cite{okada2022safety}, which underscores the disparity between the technological potential of 7T MRI and its practical accessibility for routine clinical use.

\begin{figure}[htbp]
\centering
\subfloat[1.5T]{
\label{1.5T}
\centering
\includegraphics[width=0.148\textwidth]{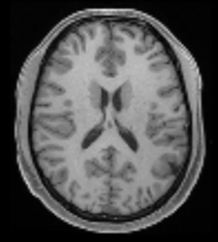}
}
\subfloat[3T]{
\label{3T}
\includegraphics[width=0.148\textwidth]{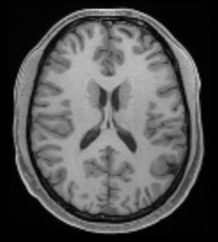}
 }
\subfloat[7T]{
\label{7T}
\centering
\includegraphics[width=0.148\textwidth]{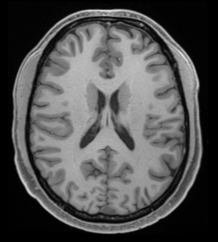}
}
\caption{An example of axial view brain MR scans at 1.5T, 3T, and 7T field strengths.}
\label{detectionknee}
\end{figure}

To address these limitations, researchers have explored various techniques for Super-Resolution (SR) MRI reconstruction. Traditionally, SR algorithms work by leveraging multiple low-resolution (LR) images of the same scene to produce a high-resolution (HR) image that surpasses the resolution limits imposed by the imaging hardware \cite{greenspan2009super}. In parallel with these traditional approaches, newer advancements have led to the development of single-image SR techniques, where HR reconstruction can be achieved from even a single LR image, broadening the scope of SR applications. Recently, several resolution enhancement methods have been proposed to reconstruct an HR image from one or more LR images. In \cite{greenspan2002mri}, Greenspan et al. reconstruct isotropic HR 3D images from 2D multi-slice MRI data. The authors address the common issue in 2D multi-slice MRI where the resolution in the slice-select direction is significantly lower than the in-plane resolution. To overcome this, they propose using multiple MRI acquisitions with sub-pixel shifts between the slices, which are then combined using an iterative SR algorithm. The method significantly improves resolution in the slice-select direction, enabling the generation of 3D isotropic images with better edge definition and superior SNR ratio efficiency compared to direct HR acquisitions. Experimental results from both phantom and human brain imaging demonstrate that the SR approach can effectively enhance spatial resolution and image quality. In \cite{peeters2004use}, Petters et al. propose an adapted Echo-Planar Imaging (EPI) acquisition protocol that acquires slice-shifted images, enabling the generation of interpolated SR images with improved resolution in the slice direction. They introduce a discontinuity-preserving regularization method to ensure the robustness of the reconstructed SR images. The method was tested on both real and synthetic fMRI datasets, where results demonstrated enhanced detectability of small activated areas compared to conventional LR data with thicker slices. The SR technique yielded higher t-values and larger activated areas in functional activation maps, highlighting its potential to significantly improve spatial resolution in fMRI while preserving image quality. In \cite{kim2009intersection}, Kim et al. present a novel approach called Sice Intersection Motion Correction (SIMC) to address the challenge of motion artefacts in 3D fetal brain MRI. Unlike traditional methods that rely on a computationally expensive slice-to-volume iterative process, the SIMC approach simplifies the alignment by directly matching intersecting slices across different orientations. The authors introduce a spatially weighted mean square intensity difference as the similarity measure for slice registration and propose an ellipsoid-based 3D spatial windowing technique to suppress the influence of maternal tissues. The authors demonstrate the effectiveness of the proposed method through both simulated and clinical fetal MRI datasets. Results show that the method can recover up to 15 mm of translation and 30 degrees of rotation, significantly improving the resolution and image quality in the final 3D reconstructions.

With the rapid advancements in deep learning technologies, the field of SR in MRI has witnessed remarkable growth. Deep learning-based SR methods leverage powerful neural network architectures to learn complex mappings between LR and HR images. In \cite{dong2014learning}, Dong et al. present a deep Convolutional Neural Network (CNN) called a Super-Resolution Convolutional Neural Network (SRCNN) for the SR task of single-image. The authors propose an end-to-end mapping between LR and HR images using a lightweight CNN architecture that avoids the traditional patch-based methods like sparse-coding. Unlike previous SR approaches that require separate dictionary learning and post-processing, SRCNN integrates the entire process into a unified optimization framework, directly learning the mapping from LR to HR images through convolutional layers. The authors demonstrate SRCNN’s superiority in terms of Peak Signal-to-Noise Ratio (PSNR) and Structural Similarity Index (SSIM), as well as its competitive runtime. In \cite{oktay2016multi}, oktay et al. introduce a novel multi-input cardiac image SR approach using CNNs to enhance the resolution of 2D multi-slice cardiac MRI. The authors propose a residual CNN model that reconstructs HR 3D volumes from 2D image stacks, improving the visualization of cardiac anatomy and the accuracy of quantitative analysis. The model leverages multiple input data from different imaging planes (short-axis and long-axis) to enhance the quality of the reconstructed images. The model also shows superior performance in cardiac segmentation and motion-tracking tasks when compared to conventional interpolation methods. In \cite{chen2018efficient}, Chen et al. introduce a 3D multi-level densely connected network (mDCSRN) and a Generative Adversarial Network (GAN) \cite{GAN} for HR image reconstruction. The authors address the challenges of recovering HR images from LR MRI scans by proposing a deep-learning model that significantly reduces computational complexity while maintaining superior image quality. The mDCSRN utilizes densely connected layers to efficiently process 3D volumetric MRI data, capturing both local and global image features for improved spatial resolution. The inclusion of GAN training further enhances the perceptual quality of the generated images, producing sharper and more realistic results. The proposed approach runs six times faster than other models while providing comparable or better visual quality, enabling a fourfold reduction in scan time without compromising resolution. In \cite{qu2020synthesized}, Qu et al. synthesize 7T MRI images from 3T MRI scans by leveraging both spatial and wavelet domains. The proposed model, named WATNet, incorporates a Wavelet-based Affine Transformation (WAT) layer to modulate spatial feature maps with information from the wavelet domain, allowing the network to capture both low-frequency tissue contrast and high-frequency anatomical details effectively. The method utilizes wavelet transformation to decompose the input 3T MRI image into different frequency components, enabling multi-scale reconstruction of 7T images.

Inspired by the above work and \cite{saharia2022image}, we propose a novel SR approach based on the Conditional Latent Diffusion Model (CLDM) \cite{rombach2022high} to generate 7T-like MRI images from LR 1.5T MRI data. The core idea leverages the power of diffusion models, which progressively refine noisy inputs into high-quality outputs through a denoising process conditioned on auxiliary information. During the training phase, individual slices from all three axes (i.e., axial, sagittal, and coronal) of 7T MRI are noised and then iteratively denoised, using their corresponding 1.5T MRI slices as conditional input alongside gradient nonlinearity correction and bias field correction guidance, which enables the model to learn to reconstruct 7T-like MRI slices by leveraging both input features and domain-specific corrections for enhanced accuracy. During inference, the model starts from Gaussian noise and progressively generates the 7T-like MRI slices by refining the signal in small steps, conditioned on the 1.5T MRI input and guidance. The process is repeated across all slices and axes, producing three separate axis-aligned volumes, which are subsequently merged through averaging to ensure spatial consistency and robustness in the final reconstructed 7T-like MRI volume. However, we did not stop there. To further enhance the practicality of such an approach, a novel progressive distillation technique is adopted. Specifically, the baseline model serves as a teacher, guiding the development of a student model that relies on intermediate feature maps generated during inference. A step count parameter $N$ is introduced to allow the student model to perform incremental SR tasks in smaller, more manageable steps, progressively approximating the teacher model’s output. Our proposed distillation strategy not only significantly reduces the parameter size and computational complexity of the student model but also eliminates its dependency on gradient nonlinearity correction and bias field correction. Additionally, the student model can accept MRI data of varying resolutions, significantly improving its adaptability and making it deployable in diverse clinical settings, combining HR reconstruction with flexibility and efficiency.

The primary contributions of this study include the following:
\begin{itemize}
\item[$\bullet$] A novel CLDM-based framework integrating gradient nonlinearity correction and bias field correction as guidance is proposed to generate 7T-like MRI volumes from real 1.5T MRI data.
\item[$\bullet$] A progressive distillation strategy is introduced to train a lightweight student model that achieves comparable performance to the teacher model while significantly improving deployability by dramatically reducing computational and resource requirements.
\item[$\bullet$] The student model supports multi-scale resolution adaptation, enabling SR tasks across different MRI resolutions without requiring additional fine-tuning or retraining.
\item[$\bullet$] The clinical evaluations conducted by experienced radiologists are performed, demonstrating that the generated higher resolution MRIs are not only visually accurate but also exhibit potential clinical significance.
\end{itemize}

\section{Proposed approach}
\begin{figure*}[htbp]
\centering 
\includegraphics[width=1\textwidth]{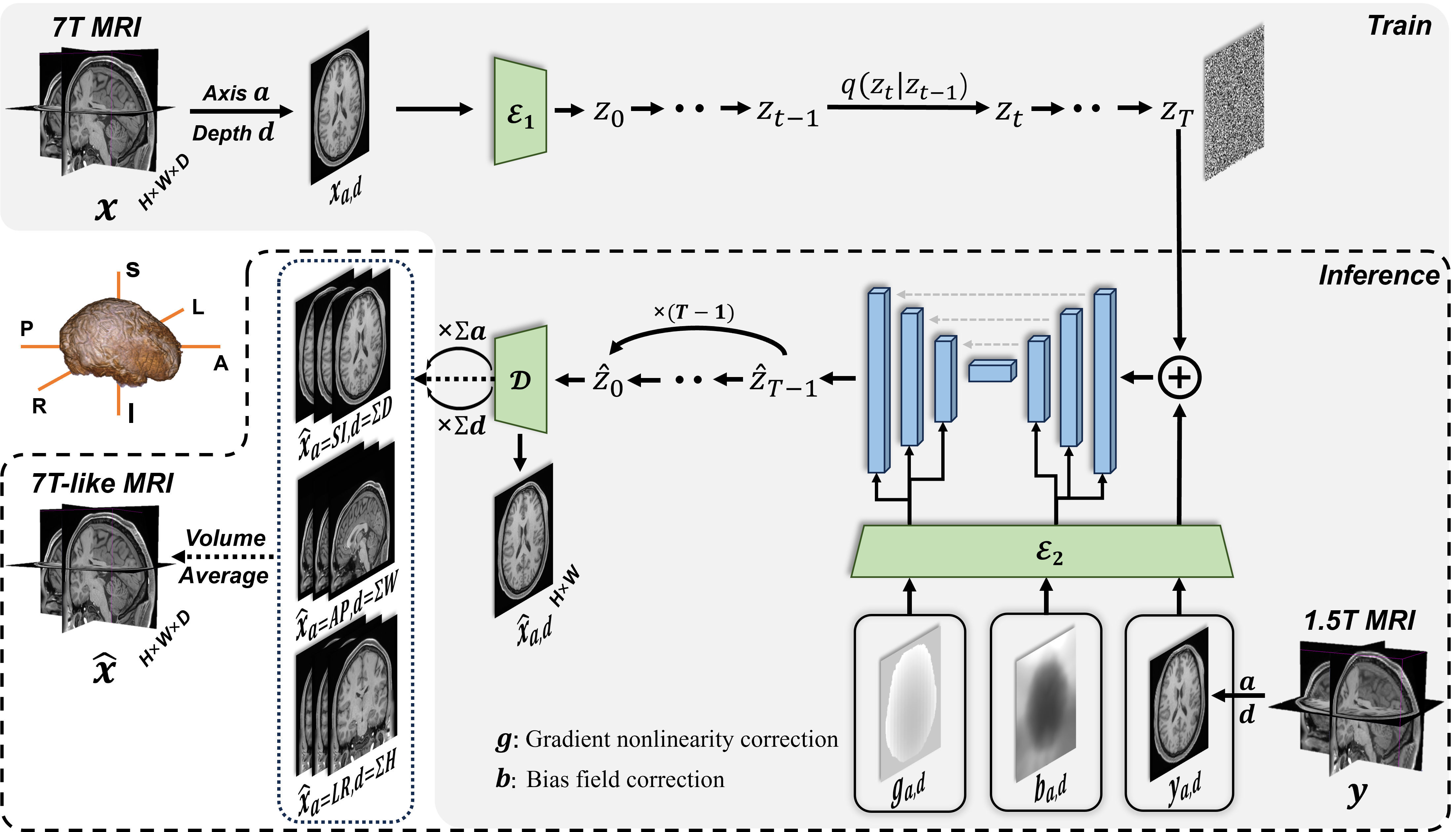}
\caption{The architecture of the baseline (teacher) model begins by extracting a slice at position $d$ along axis $a$ from a real MRI sequence $x$, denoted as $x_{a,d}$. This slice is encoded by the encoder $\mathcal{E}_1$, producing an initial latent representation $z_0$ that is progressively noised to a noised latent representation $z_T$. The corresponding slice $y_{a,d}$ from a 1.5T MRI sequence $y$, positioned at the same axis and depth, is used as a conditional input. This slice is encoded by a second encoder $\mathcal{E}_2$, and the output is concatenated with $z_T$ to form the initial input for the downsampling network. Throughout the denoising process, the bias field correction $b$ and the gradient nonlinearity correction $g$ are embedded at specific stages of the downsampling and upsampling phases, respectively, to provide guidance, culminating in the denoised latent $\hat{z}_0$. The decoder $\mathcal{D}$ then reconstructs the corresponding MRI slice $\hat{x}_{a,d}$ from $\hat{z}_0$. During inference, the process begins with $z_T$ and $y$, and is repeated for each slice at all positions $(\sum a, \sum d)$. The final 7T-like MRI sequence $\hat{x}$ is constructed by stacking all generated MRI slices in sequential order along each axis, followed by volume averaging.}
\label{flowchart}
\end{figure*}

\subsection{Baseline (Teacher) model}
\label{teacher_model}
As presented in Fig. \ref{flowchart}, the proposed teacher model consists of three main modules: an Auto-Encoder (AE) module, a CLDM module, and a guidance module. $x_{a, d}$ is a slice of the original knee MRI sequence $x$ at the depth position $d$ in the axis $a$ with the size of $H \times W \times D$. $\mathcal{E}_1$ and $\mathcal{D}$ are encoder and decoder of the AutoencoderKL \cite{kingma2013auto}. Unlike traditional autoencoders that map input data to a deterministic latent space, the AutoencoderKL introduces a probabilistic approach by encoding the input into a Gaussian distribution over the latent space. During training, the Kullback-Leibler (KL) divergence is used as a regularization term to ensure that the learned latent space closely follows a standard normal distribution, promoting smoother and more continuous transitions between latent representations. Specifically, the encoder operates at a resolution of 256$\times$256$\times$3 and begins with a convolutional layer configured with 128 channels and then proceeds through a series of stages, each with 2 residual blocks. The channel dimensions are multiplied according to the sequence 1, 2, 4, and 4. The decoder mirrors the encoder’s structure, reversing the feature extraction process to reconstruct high-quality outputs. Our design deliberately excludes dropout layers, prioritizing the retention of all learned features without stochastic regularization during training. For the AE, the reconstruction loss $\mathcal{L}_ {rec}$ is as:

\begin{equation}
\label{encoder-decoder-loss}
\mathcal{L}_ {rec} = ||(x_{a,d}, \mathcal{D}(\mathcal{E}_1(x_{a,d}))||_{2} ^ {2}
\end{equation}

After training the AE, for a given input slice $x_{a,d}$, it is encoded by the parameter-frozen encoder $\mathcal{E}_1$ into the latent representation $z_0$. The forward diffusion process begins with the original latent representation $z_0 \sim q(z_0)$, where Gaussian noise $n$ is gradually introduced in a controlled manner over a total of $T$ steps. At each diffusion step $t$, the noised latent data $z_{t-1}$ from the previous step undergoes a transformation to generate $z_t$, the noised latent data at the current step. The transformation follows the forward diffusion rule defined as:

\begin{equation}
q(z_t\mid z_{t-1})=\mathcal{N} (z_t;\sqrt[]{1-\beta_t}z_{t-1},\beta_t\mathbf{I})
\end{equation}
where $\beta_t$ represents the variance used for each step in the range of $[0,1]$, and $\mathbf{I}$ represents the identity matrix. The entire diffusion process constitutes a Markov chain \cite{norris1998markov}:
\begin{equation}
q(z_{T}\mid z_0)=\prod_{t=1}^{T} q(z_t\mid z_{t-1})
\end{equation}

Through the reparameterization technique \cite{kingma2015variational}, the noised latent code $z_T$ is calculated as:
\begin{equation}
\label{noise}
z_T = \sqrt{\bar{\alpha}_T}\mathcal{E}_1(x_{a,d}) + \sqrt{1-\bar{\alpha}_T}\epsilon
\end{equation}
where $\bar{\alpha}_T = \prod_{t=1}^T \alpha_t$ represents the cumulative product of time-dependent coefficients $\alpha_t$ over $T$ diffusion steps, which represents the overall attenuation factor of the original signal across all steps. $\alpha_t$ is a noise scheduling parameter at step $t$, controlling the proportion of the original signal preserved versus the noise added. $\epsilon \sim \mathcal{N}(0, I)$ denotes Gaussian noise sampled from a standard normal distribution, added to the latent code.

Conditional input $y$ is a corresponding 1.5T MRI volume. Similar to the target slice $x_{a,d}$, $y_{a,d}$ refers to the slice located at a depth $d$ along the axis $a$ within $y$. Each $y_{a,d}$ is encoded by the encoder $\mathcal{E}_2$, generating a feature representation that is concatenated with the noised latent code $z_T$ as the input of the U-Net, denoted as $\mathcal{U}$, serves as the beginning of the denoise process. The employed 2D U-Net architecture adopts a multi-scale hierarchical structure comprising a time embedding module, an encoder module, a middle module, and a decoder module. This UNet model takes 8-channel input and produces 4-channel output, built upon a base channel configuration of 320. The encoder module begins with a 2D convolutional layer featuring a kernel size of 3 and padding of 1, followed by a series of residual blocks with progressively increasing feature dimensions. Each residual block includes multiple convolutional layers, GroupNorm \cite{wu2018group} for normalization and the Swish activation function \cite{ramachandran2017searching} for non-linearity. Two residual blocks are employed per downsampling stage, with attention mechanisms activated at resolutions of 4, 2, and 1. The channel count increases across network layers following a multiplier sequence of 1, 2, 4, and 4. The decoder mirrors this structure, utilizing the same residual blocks to progressively upsample feature maps, reducing the number of channels in reverse order while merging them with corresponding encoder feature maps through skip connections. Additionally, the model integrates 8 attention heads and employs a spatial transformer block with a depth of 1, conditioned on a context dimension of 768.

Next, the 7T bias field correction, denoted as $b$, is embedded into the downsampling stage of the model to compensate for intensity inhomogeneities, mitigating variations caused by non-uniform magnetic fields and scanner imperfections. Simultaneously, the 7T gradient nonlinearity correction, denoted as $g$, is integrated into the upsampling stage to rectify spatial distortions introduced by gradient coil imperfections, ensuring that the spatial relationships within the data are accurately reconstructed, preserving anatomical accuracy. The overall goal function $\mathcal{L}_{diff}$ for the model $\mathcal{U}$, combining these correction mechanisms, is computed as follows:

\begin{equation}
\label{DDIMM}
\mathop{\min}_{\theta} \mathbb{E}_ {z \sim \varepsilon_1(x_{a,d}), t, \epsilon \sim \mathcal{N}(0,\mathbf{I})} || \epsilon -  \epsilon_\theta \{z_t,t, [z_T, y, (g,b)]\}  ||_{2} ^ {2}
\end{equation}

During the denoising process, the latent variable $z_{t-1}$ is sampled using the Denoising Diffusion Implicit Model (DDIM) \cite{song2020denoising}, which enables deterministic or stochastic generation depending on the chosen schedule. DDIM refines the sampling process by approximating the reverse diffusion trajectory more efficiently than traditional diffusion models. Specifically, it leverages a non-Markovian process that skips some diffusion steps, allowing for faster inference without compromising the quality of the generated images. The sampling of $z_{t-1}$ is computed as follows:
\begin{equation}
z_{t-1} = \sqrt{\bar{\alpha}_{t-1}} \left(\frac{z_t - \sqrt{1 - \bar{\alpha}_t} \mathbf{\epsilon}_\theta}{\sqrt{\bar{\alpha}_t}}\right) + \sqrt{1 - \bar{\alpha}_{t-1}} \mathbf{\epsilon}_\theta
\end{equation}
where $\bar{\alpha}_t$ and $\bar{\alpha}_{t-1}$ represent the cumulative product of the noise scheduling coefficients at steps $t$ and $t-1$, respectively. The term $\mathbf{\epsilon}_\theta$ is the model-predicted noise, which enables precise reconstruction by iteratively refining $z_t$ through the reverse process.

For each reconstructed slice along a specific axis $a$ and at depth $d$, denoted as $\hat{x}_{a,d}$, the HR details are reconstructed using a decoder $\mathcal{D}$ by taking the denoised latent representation as input and maps it back into the image space. During the inference phase, this process is repeated along each axis (i.e., coronal, sagittal, and axial) to generate HR slices at all depths for each respective axis. The complete set of reconstructed slices across all axes can be represented as $\bigcup_{a}^{axes} \{\hat{x}_{a,1}, \hat{x}_{a,2}, \dots, \hat{x}_{a, d}\}$.  After generating slices for all axes, the slices are combined to form a consistent 3D volume by stacking slices along their respective axes and applying volumetric averaging. The final reconstructed super-resolved 7T-like MRI volume, $\hat{x}$, has the same dimensions as the input MRI volume, $H \times W \times D$. For further clarity, Algorithm \ref{algorithm1} and Algorithm \ref{algorithm2} detail the training and inference process of the proposed approach, respectively.

\begin{algorithm}
\caption{Training process of the teacher model}
\textbf{Input:} $t$, $x$, $y$, $g$, $b$, $\mathcal{E}_2$ and all initial parameters\\
\textbf{Output:} $\mathcal{E}_1$, $\mathcal{D}$, $\mathcal{U}$ and learned parameters\\
\While{\textnormal{\textit{not converged}}}{
\For{$a$ in $\{Axial, Sagittal, Coronal\}$}{
\For{$d$ in len(a)}{
\textbf{Get} $x_{a,d} \leftarrow x_a[d]$\\
\textbf{Compute} $\mathcal{L}_{rec}(x_{a,d}, \mathcal{D}(\mathcal{E}_1({x_{a,d}})))$ \Comment{Eq. \ref{encoder-decoder-loss}}\\
\textbf{Update} Parameters of $\mathcal{E}_1$ and $\mathcal{D}$\\}
}}
\textbf{Freeze} Parameters of $\mathcal{E}_1$ and $\mathcal{D}$\\
\While{\textnormal{\textit{not converged}}}{
\For{$a$ in $\{Axial, Sagittal, Coronal\}$}{
\For{$d$ in len(a)}{
\textbf{Get} $x_{a,d}, y_{a,d} \leftarrow x_a[d], y_a[d]$\\
\textbf{Get} $\epsilon,  z_T \leftarrow noise(\mathcal{E}_1({x_{a,d}}), t)$ \Comment{Eq. \ref{noise}}\\
\textbf{Get} $\epsilon_\theta, \hat{z}_{a,d} \leftarrow \mathcal{U}(z_T, \mathcal{E}_2(y_{a,d}), t, g, b)$\\
\textbf{Compute} $\mathcal{L}_{diff}(\epsilon, \epsilon_\theta)$ \Comment{Eq. \ref{DDIMM}}\\
\textbf{Compute} $\mathcal{L}_{rec}(x_{a,d}, \mathcal{D}(\hat{z}_{a,d}))$ \Comment{Eq. \ref{encoder-decoder-loss}}\\
\textbf{Update} Parameters of $\mathcal{U}$\\}
}}
\label{algorithm1}
\end{algorithm}

\begin{algorithm}
\caption{Inference process of the teacher model}
\textbf{Input:}  $t$, $y$, $g$, $b$, $\mathcal{E}_2$, $\mathcal{D}$ and $\mathcal{U}$\\
\textbf{Output:} $\hat{x}$\\
\textbf{Get} $\epsilon \leftarrow \mathcal{N}(0, \mathbf{I})$\\
\For{$a$ in $\{Axial, Sagittal, Coronal\}$}{
\For{$d$ in len(a)}{
\textbf{Get} $y_{a,d} \leftarrow y_a[d]$\\
\textbf{Get} $\hat{z}_{a,d} \leftarrow \mathcal{U}(\epsilon, \mathcal{E}_2(y_{a,d}), t, g, b)$\\
\textbf{Get} $\hat{x}_{a,d} \leftarrow \mathcal{D}(\hat{z}_{a,d})$\\}
\textbf{Get} $\hat{x}_a \leftarrow \bigcup_{d}^{len(a)} \hat{x}_{a,d}$\\}
\textbf{Get} $\hat{x} \leftarrow \bigcup_{a}^{axes} \hat{x}_{a}$\\
\label{algorithm2}
\end{algorithm}

\subsection{Progressive distillation strategy}
In \cite{salimans2022progressive}, Tim et al. introduced a distillation method that progressively halved sampling steps through multiple distillation cycles, and achieved similar quality with significantly fewer steps, accelerating the sampling process of diffusion models. On the other hand, in \cite{gao2023implicit}, Gao et al. introduced a scale-adaptive conditioning mechanism, allowing the model to adjust the ratio of low-resolution information to generated high-detail features dynamically. This approach leverages a diffusion process to iteratively refine outputs through an implicit image function, enhancing visual detail and realism without fixed magnification constraints. 

\begin{figure}[htbp]
\centering
\includegraphics[width=0.5\textwidth]{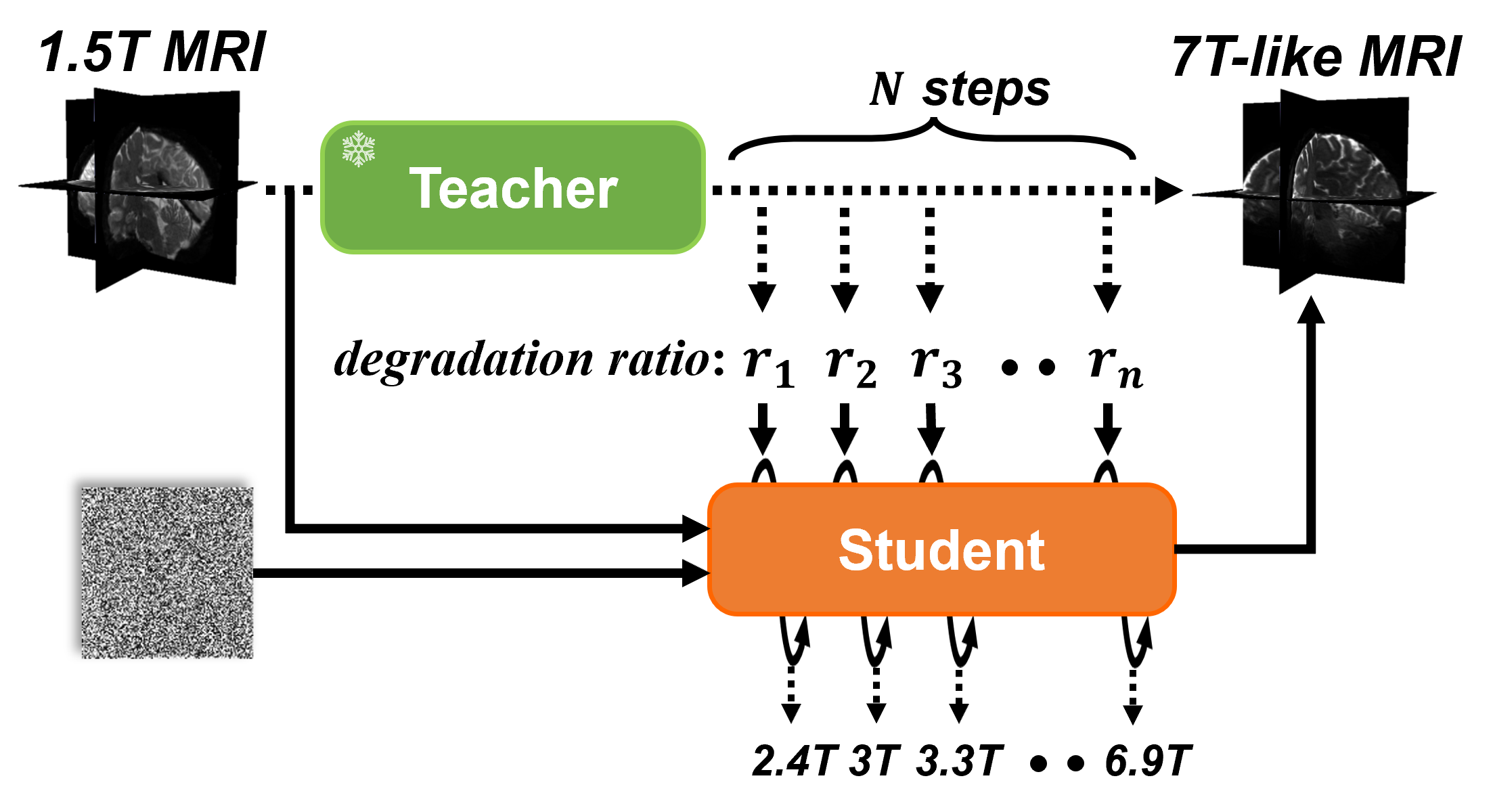}
\caption{The illustration showcases the progressive knowledge distillation process. The Teacher is highlighted, operating as a high-capacity baseline model capable of generating high-quality 7T-like MRI outputs from 1.5T inputs. The Student is depicted as a lightweight architecture designed to learn from the teacher model’s outputs. The arrows connecting the teacher and student models emphasize the flow of knowledge and guidance during training. The process leverages progressive distillation, where the student model incrementally refines its latent representations by matching intermediate targets provided by the teacher model.}
\label{flowchart_global}
\end{figure}

Inspired by the above work, as depicted in Fig. \ref{flowchart_global}, we propose a novel progressive distillation strategy to refine the SR process by enabling the more streamlined student model to iteratively approach the target 7T MRI resolution from an initial 1.5T MRI over multiple steps under the guidance of the large and high-capacity teacher model. A key innovation is the introduction of a step count parameter, $N$, which represents the total number of stages in the resolution refinement process. By adjusting $N$, the granularity of the intermediate resolution targets can be controlled, providing flexibility in shaping the learning trajectory of the student model. For each given step count $N$, each intermediate target resolution is associated with a degradation ratio relative to the real 7T resolution. This ratio is used to degrade the 7T-like outputs of the teacher model, serving as sub-targets for the student model at each training stage. To achieve this, a Gaussian blur is applied to create the intermediate target $I_{\text{target}}^{(n)}$ for the student model at stage $n$. The process is formulated as:

\begin{equation}
\label{degrade}
I_{\text{target}}^{(n)} = O_\text{7T} * \mathcal{G}(0, k \cdot r_n)
\end{equation}
where $O_{\text{7T}}$ is the 7T-like output from the teacher model, $r_n$ is the degradation ratio at stage $n$, controlling the level of degradation. $k$ is a proportional constant to adjust degradation strength, $\mathcal{G}(0, k \cdot r_n)$ is a Gaussian kernel with standard deviation $k \cdot r_n$, $*$ denotes convolution, applying the degradation to create the intermediate target resolution without altering the image size.

The student model then minimizes the following objective function during each stage $n$:

\begin{equation}
\mathop{\min}_{\theta} \mathbb{E}_ {z \sim \varepsilon_1(I_{\text{target}}^{(n)}), t, \epsilon \sim \mathcal{N}(0,\mathbf{I})} \big\| \epsilon -  \epsilon_\theta [z_t, t, (z_T, y)] \big\|_{2}^2
\end{equation}
where $\epsilon_\theta$ represents the student model's predicted noise, $z_t$ is the noised latent variable at step $t$, and $z_T$ and $y$ are noised degraded teacher output and the current resolution input, respectively.

For optimal guidance, a total of 199 feature maps, denoted as $\mathcal{F}_{\text{teacher}}$, are extracted from the teacher model across its inference steps. Unlike the student model, which achieves the 1.5T to 7T progressive SR task through a series of intermediate subgoals, the teacher model performs the task in a relatively single-step manner without explicit stage-by-stage progression. As a result, these teacher feature maps are not partitioned to align with any predefined subgoals. Instead, they function as a global reference, providing the student model with holistic guidance throughout its training, which allows the student model to progressively refine its outputs and approximate the 7T-like target resolution by leveraging the broader contextual and structural information encoded in the teacher’s feature representations. In this setup, the feature alignment loss serves as the distillation loss, which is formulated as follows:
\begin{equation}
\mathcal{L}_{\text{distillation}} = \frac{1}{N} \frac{1}{K}\sum_{n}^{N} \sum_{k}^{K}  r_n \big\|(O_{\text{student}}^{(n)} - \mathcal{F}_{\text{teacher}}^{(k)})\big\|_2^2
\end{equation}
where $r_n$ is a degradation factor specific to each stage $n$. $O_{\text{student}}^{(n)}$ denotes the output of the student model at stage $n$. $K$ is the total number of feature maps used from the teacher model, which is set to 199 in this study. $\mathcal{F}_{\text{teacher}}^{(k)}$ denotes the $k$-th feature map of the teacher model.

The student model then matches these intermediate targets by minimizing the distillation loss, progressively refining its output until it closely aligns with the teacher model's HR output. Beyond the distillation part, the overall training and inference strategy remains consistent with that of the teacher model. Such a lighter student diffusion model structure is designed with 2 input channels, 1 output channel, and a scalable inner channel size starting at 64, incorporating channel multipliers of 1, 2, 2, 4, 4, 8, 8 and 16. The inclusion of two residual blocks per stage stabilizes learning and improves feature retention, while an attention layer at a resolution of 16. To further streamline the model, the encoder $\mathcal{E}_2$ is removed, and the 1.5T MRI slice used as a conditional input is concatenated directly with the noised intermediate target at each stage. Additionally, both the bias field $b$ and gradient nonlinearity correction $g$ guidance are removed, further enhancing the model’s deployability by removing dependencies on external corrections. These modifications significantly reduce the computational complexity and memory requirements of the student model, making it more suitable for clinical practice applications where hardware resources may be limited.

\section{Experimental settings}

\subsection{Employed database}
The Human Connectome Project (HCP) maps the healthy human connectome by collecting neuroimaging and behavioural data on 1,200 normal young adults, aged 22-35. The project was carried out in two phases by a consortium of over 100 investigators and staff at 10 institutions. In Phase I, data acquisition and analysis methods were optimized, including refinements to pulse sequences and key preprocessing steps. In Phase II, neuroimaging and behavioural data were acquired from 1,200 healthy adults recruited from 300 families of twins and their non-twin siblings. To obtain brain connectivity maps of the highest quality, HCP employed cutting-edge MR hardware, including 3T and 7T MR scanners and customized head coils. 3T MR was completed on a single Siemens Skyra Connectom scanner for 1113 subjects at Washington University, and 7T MR was collected on a subset of 184 subjects on a Siemens Magnetom scanner at UMinn.

\subsection{Data preprocessing}
\label{data_preprocessing}
After normalization and standardization, each 1.5T and 7T MRI volume with a respective voxel size of 2mm $\times$ 2mm $\times$ 2mm and 0.7mm $\times$ 0.7mm $\times$ 0.7mm are resampled to a consistent size of $260 \times 310 \times 260$ voxels, where each slice is reformatted to a spatial resolution of $256 \times 256 \times 3$ pixels. To meet the requirements of the pre-trained diffusion model with a channel size of 3, the grayscale images are duplicated three times and then concatenated along the channel dimension. The resulting dataset consists of 184 pairs of T1-weighted (T1w) and T2-weighted (T2w) MRI volumes at both 1.5T and 7T resolutions, respectively. 

\subsection{Experimental details}
\subsubsection{Teacher model} The diffusion module utilized pre-trained weights from Zero-123 \cite{Liu_2023_ICCV}. The pre-trained encoder $\mathcal{E}_2$ is sourced from \cite{pmlr-v139}, where the integrated encoder is sourced from the ViT-L/32 \cite{dosovitskiy2020image}. During the training, a base learning rate of 1e-06 was applied, with a lambda linear scheduler incorporating a warm-up period every 100 steps. The batch size was set to 64, with $T$ configured to 5,000 steps, and an Exponential Moving Average (EMA) \cite{hunter1986exponentially} of 0.99 was used for the diffusion sampling process. The scale factor of the latent space was set as 0.2. The AdamW optimizer \cite{loshchilov2019} was utilized for training over 5,000 epochs. 
\subsubsection{Student model} The diffusion module was trained from scratch with a base learning rate of 1e-04. The batch size was set to 192, with $T$ configured to 2,000 steps. All other configurations and training strategies were kept consistent with those used for the teacher model.

We implemented our global approach using PyTorch v1.12.1 \cite{pytorch} on Nvidia A100 80 GB graphics cards.

\section{Results and discussion}
\subsection{Ablation study for the guidance modules}
In our baseline (teacher) model, bias field correction $b$ and gradient nonlinearity correction $g$ are specifically designed to address inherent artefacts and distortions that can arise in high-field MR scans. To investigate the importance of the guidance modules, an ablation study is conducted to evaluate how the inclusion or exclusion of these modules affects the model's performance on 7T-like outputs. The results of the ablation study are presented in Table \ref{ablation_study}, where LPIPS scores \cite{zhang2018unreasonable} are used to quantify the perceptual similarity between the generated outputs and the 7T ground truth. Lower LPIPS scores indicate higher perceptual similarity and better alignment with the ground truth. When both $b$ and $g$ are employed, the student model achieves the best performance with an LPIPS score of 0.0353, significantly outperforming configurations that exclude either or both guidance modules, which underscores the synergistic effect of combining $b$ and $g$ in addressing both intensity inhomogeneities and spatial distortions. For context, the LPIPS difference between the original 1.5T MRI and the 7T ground truth is 0.0896, serving as a baseline reference. The results demonstrate that while partial guidance improves performance compared to no guidance, the combined use of bias field correction and gradient nonlinearity correction is crucial for minimizing perceptual discrepancies and achieving optimal reconstruction quality.

\begin{table}[htbp]
\centering
\caption{Analysis of impact of guidance modules}
\label{ablation_study}
\setlength{\tabcolsep}{4mm} 
\begin{tabular}{c|cc}
\toprule
\diagbox[width=18.6em]{$g$}{\scriptsize LPIPS}{$b$} 
& with $b$ & without $b$ \\ 
\hline
\\[-1.5ex] 
with $g$ & \bf 0.0353 & 0.0437 \\ 
without $g$ & 0.0518 & 0.0634 \\ 
\midrule
real 1.5T vs real 7T (context reference) & \multicolumn{2}{c}{0.0896}\\
\bottomrule
\end{tabular}
\end{table}

\subsection{Performance comparisons with state-of-the-art models}
The comparison is divided into two main parts: Qualitative visualization analysis and quantitative metric-based analysis.

\subsubsection{Qualitative visualization analysis}
Table \ref{visualization} provides a visual comparison of 7T-like MRI generated by various State-Of-The-Art (SOTA) SR methods using 1.5T MRI as input. The compared methods include ESRGAN \cite{wang2018esrgan}, SR3 \cite{saharia2022image}, and our proposed baseline (teacher) model. To facilitate a focused and intuitive assessment, the cerebellar region, known for its intricate structure and HR requirements, is specifically selected for comparison. To comprehensively evaluate the methods, T1w and T2w slices from both the axial and coronal planes are displayed in an alternating interleaved format, which highlights each method's effectiveness across distinct anatomical views and tissue contrasts, allowing for a clearer assessment of their ability to capture fine structural details. The comparison reveals that ESRGAN enhances the resolution but introduces significant artifacts, where unnatural patterns distort the anatomical structures. While it provides some sharpening, the output deviates substantially from the ground truth. SR3, on the other hand, prioritizes smoothness, leading to blurred outputs that fail to recover fine details and textures, which results in a loss of critical anatomical information. In contrast, the proposed teacher model excels in preserving fine structural details and reconstructs intricate anatomical features without introducing artefacts or excessive blurring, demonstrating superior performance compared to ESRGAN and SR3 in both visual quality and alignment with the 7T ground truth. Additionally, the comparison highlights a consistent discrepancy in SR performance between T1w and T2w images. T2w images generally exhibit inferior reconstruction quality, which can be attributed to their inherently lower SNR and higher susceptibility to magnetic field inhomogeneities. These characteristics pose significant challenges for SR models, making accurate reconstruction more difficult. Despite these challenges, our proposed model consistently outperforms ESRGAN and SR3, delivering superior results for both T1w and T2w imaging.
\begin{table*}[htbp]
\centering
\caption{Qualitative visualization comparison with SOTA models}
\label{visualization}
\begin{threeparttable}
\setlength{\tabcolsep}{0.2mm}
\begin{tabular}{cccccc}
\toprule
 & Input (1.5T) & Ground Truth (7T)& ESRGAN & SR3  & Ours (Teacher)\\
\midrule
\rotatebox[origin=c]{90}{T1w}&\begin{minipage}[b]{0.4\columnwidth}\centering \raisebox{-.5\height}{\includegraphics[width=\linewidth]{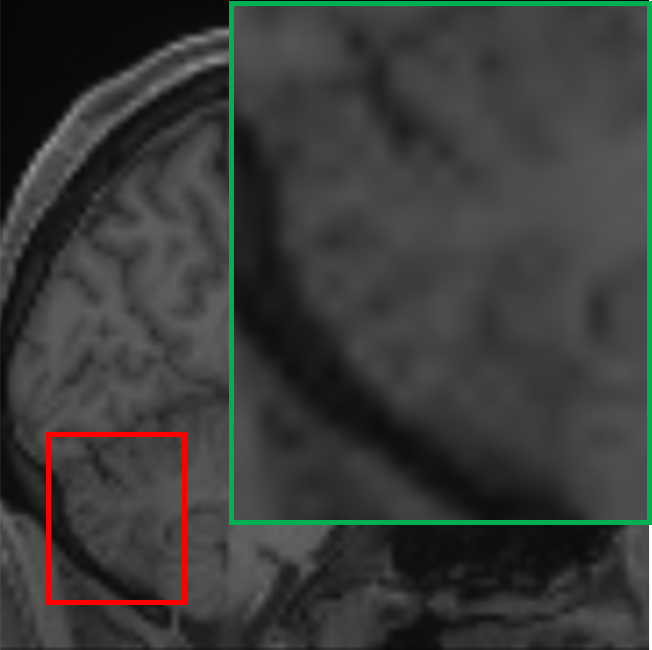}}\end{minipage}&\begin{minipage}[b]{0.4\columnwidth}\centering \raisebox{-.5\height}{\includegraphics[width=\linewidth]{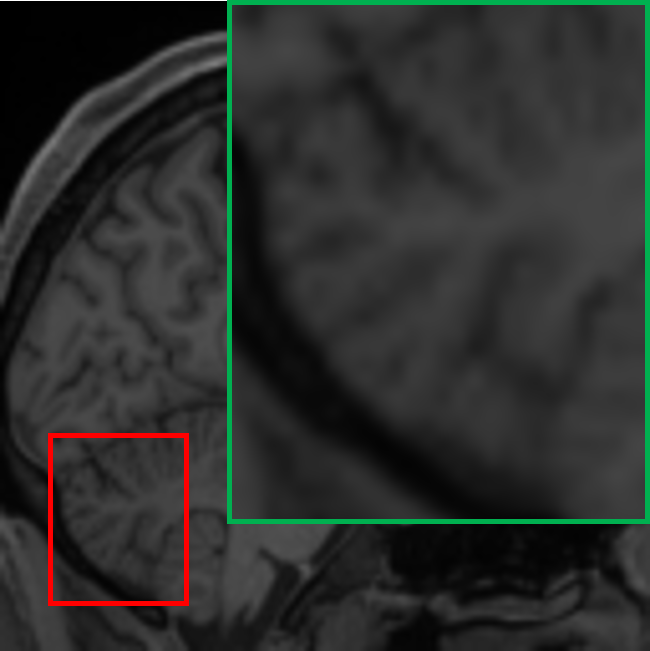}}\end{minipage}&\begin{minipage}[b]{0.4\columnwidth}\centering \raisebox{-.5\height}{\includegraphics[width=\linewidth]{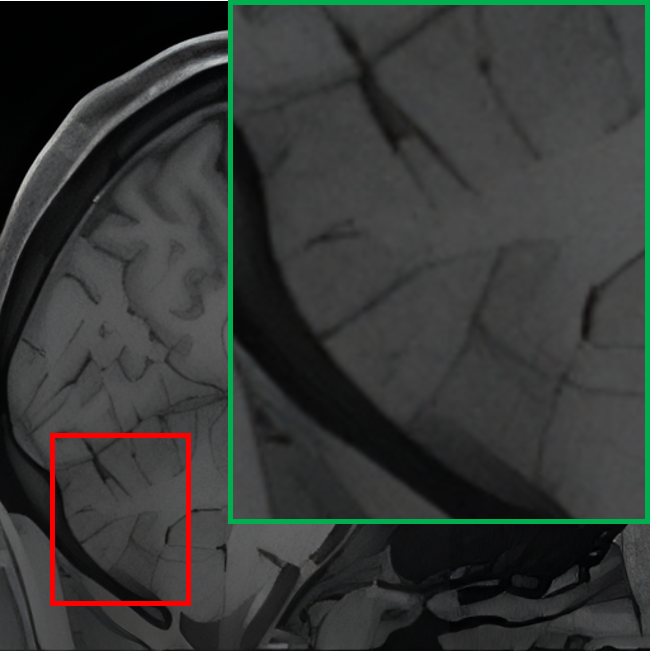}}\end{minipage}&\begin{minipage}[b]{0.4\columnwidth}\centering \raisebox{-.5\height}{\includegraphics[width=\linewidth]{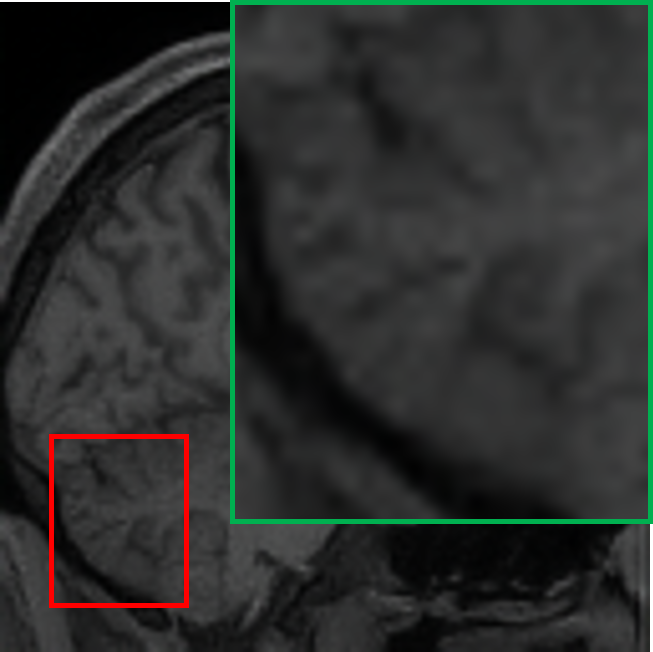}}\end{minipage}&\begin{minipage}[b]{0.4\columnwidth}\centering \raisebox{-.5\height}{\includegraphics[width=\linewidth]{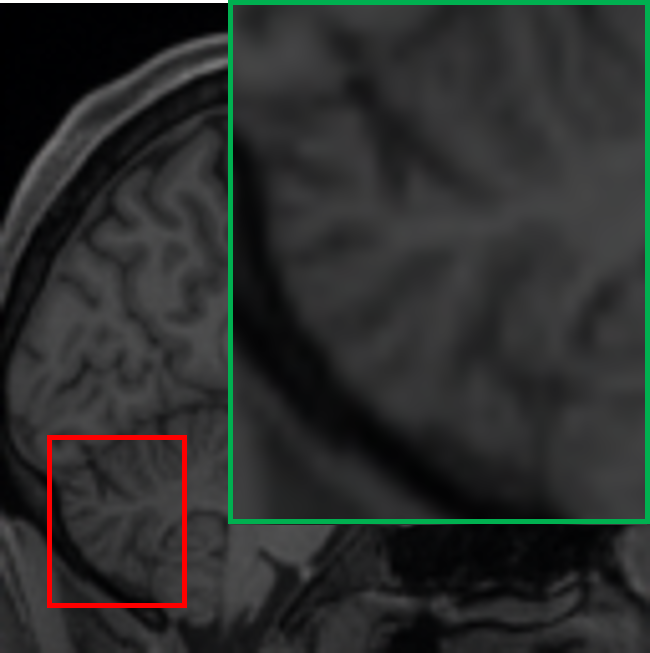}}\end{minipage}\\
\midrule
\rotatebox[origin=c]{90}{T2w}&\begin{minipage}[b]{0.4\columnwidth}\centering \raisebox{-.5\height}{\includegraphics[width=\linewidth]{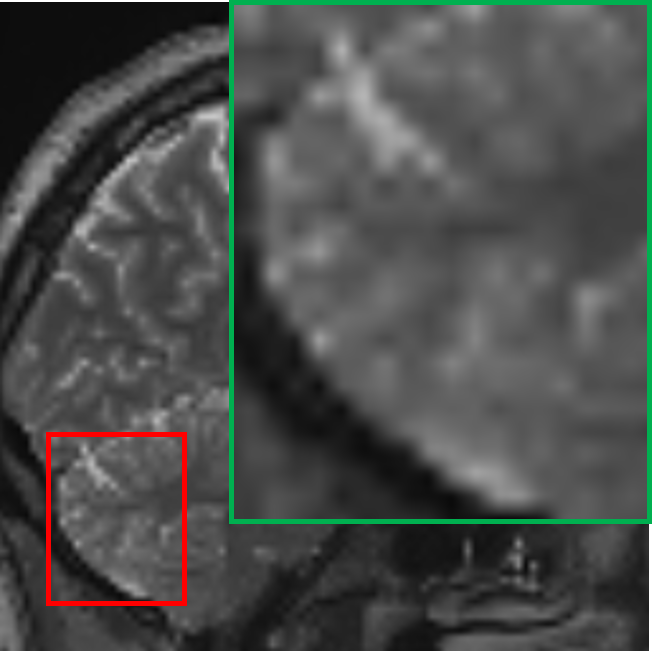}}\end{minipage}&\begin{minipage}[b]{0.4\columnwidth}\centering \raisebox{-.5\height}{\includegraphics[width=\linewidth]{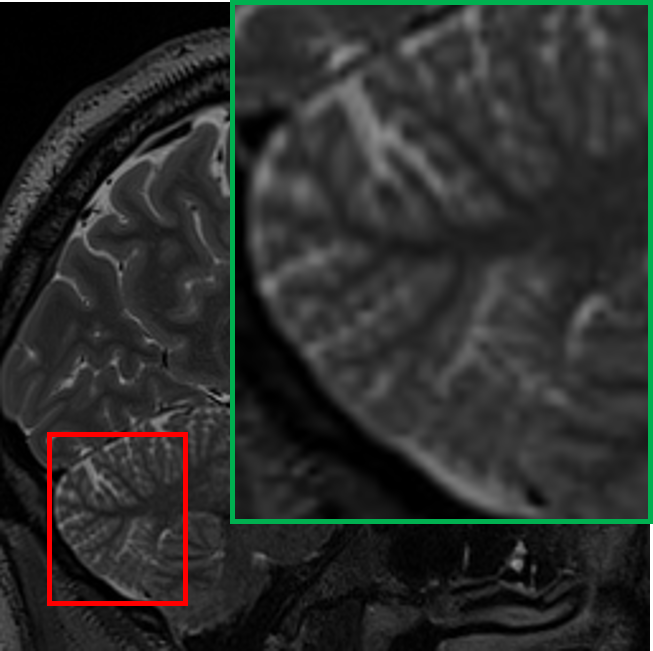}}\end{minipage}&\begin{minipage}[b]{0.4\columnwidth}\centering \raisebox{-.5\height}{\includegraphics[width=\linewidth]{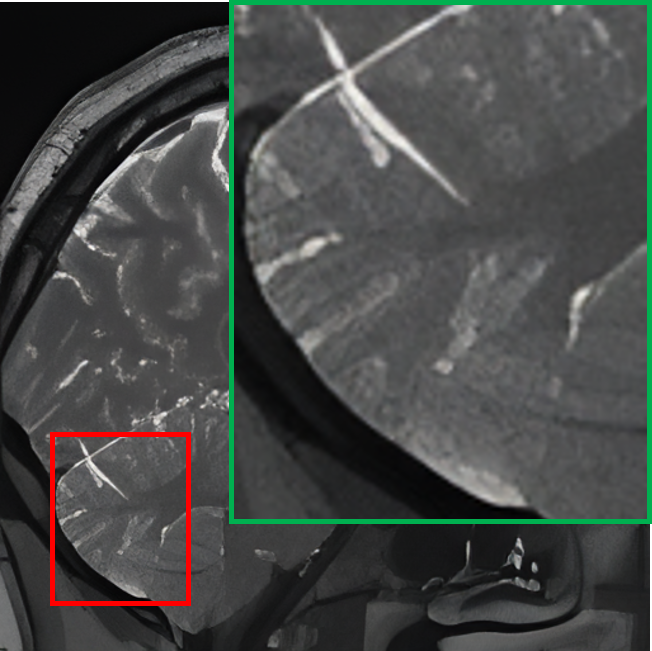}}\end{minipage}&\begin{minipage}[b]{0.4\columnwidth}\centering \raisebox{-.5\height}{\includegraphics[width=\linewidth]{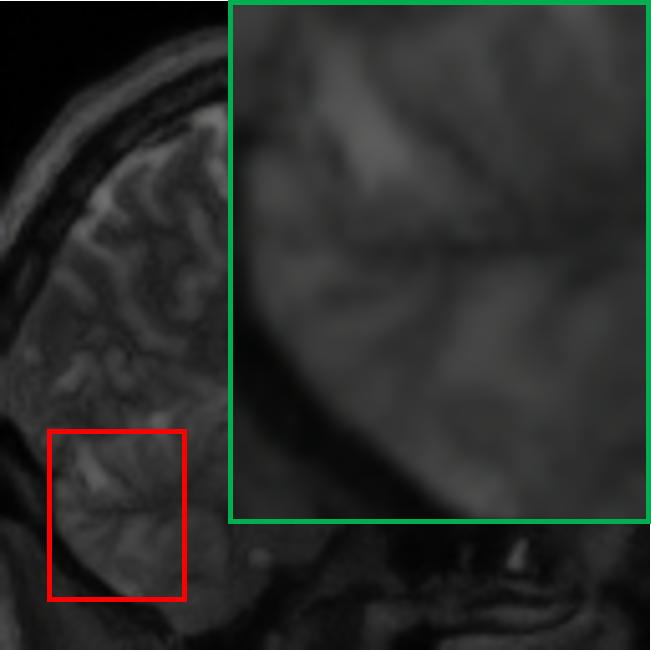}}\end{minipage}&\begin{minipage}[b]{0.4\columnwidth}\centering \raisebox{-.5\height}{\includegraphics[width=\linewidth]{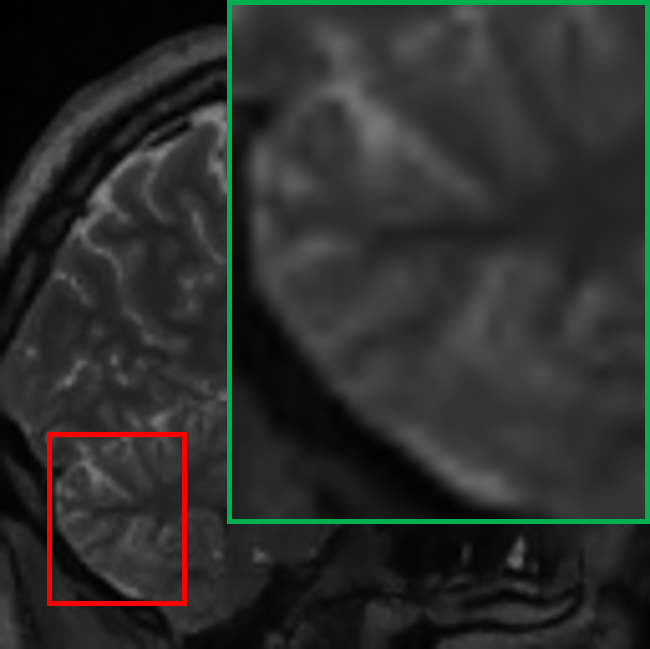}}\end{minipage}\\
\midrule
\rotatebox[origin=c]{90}{T1w}&\begin{minipage}[b]{0.4\columnwidth}\centering \raisebox{-.5\height}{\includegraphics[width=\linewidth]{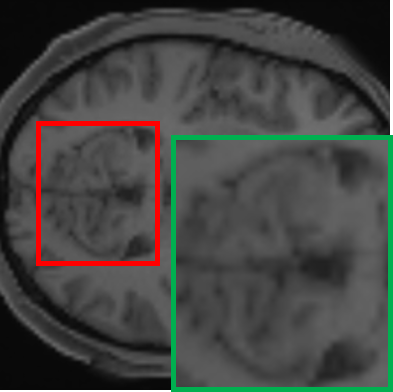}}\end{minipage}&\begin{minipage}[b]{0.4\columnwidth}\centering \raisebox{-.5\height}{\includegraphics[width=\linewidth]{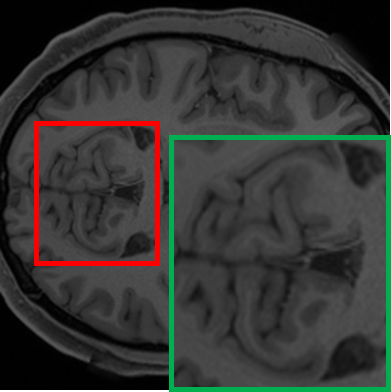}}\end{minipage}&\begin{minipage}[b]{0.4\columnwidth}\centering \raisebox{-.5\height}{\includegraphics[width=\linewidth]{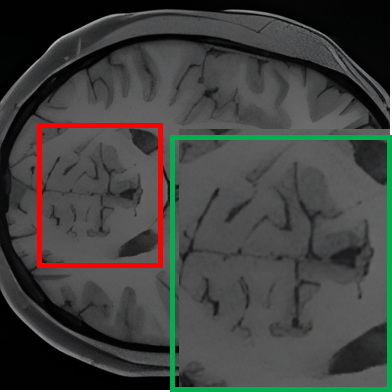}}\end{minipage}&\begin{minipage}[b]{0.4\columnwidth}\centering \raisebox{-.5\height}{\includegraphics[width=\linewidth]{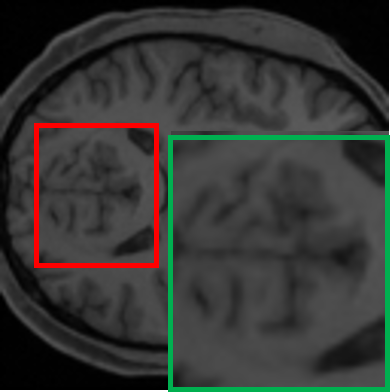}}\end{minipage}&\begin{minipage}[b]{0.4\columnwidth}\centering \raisebox{-.5\height}{\includegraphics[width=\linewidth]{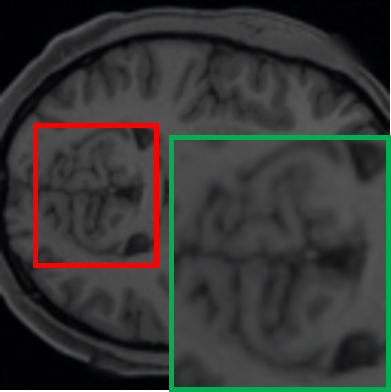}}\end{minipage}\\
\midrule
\rotatebox[origin=c]{90}{T2w}&\begin{minipage}[b]{0.4\columnwidth}\centering \raisebox{-.5\height}{\includegraphics[width=\linewidth]{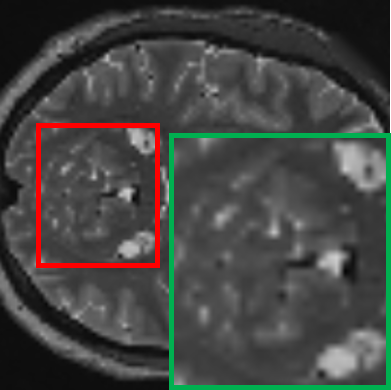}}\end{minipage}&\begin{minipage}[b]{0.4\columnwidth}\centering \raisebox{-.5\height}{\includegraphics[width=\linewidth]{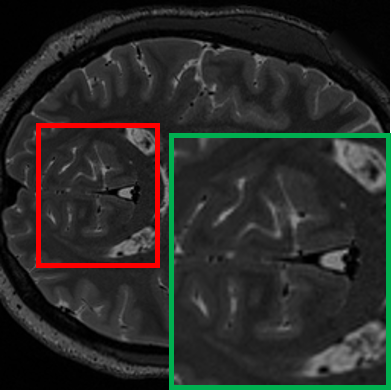}}\end{minipage}&\begin{minipage}[b]{0.4\columnwidth}\centering \raisebox{-.5\height}{\includegraphics[width=\linewidth]{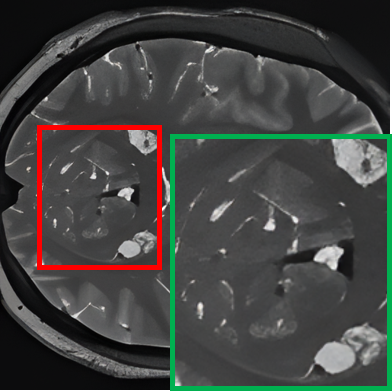}}\end{minipage}&\begin{minipage}[b]{0.4\columnwidth}\centering \raisebox{-.5\height}{\includegraphics[width=\linewidth]{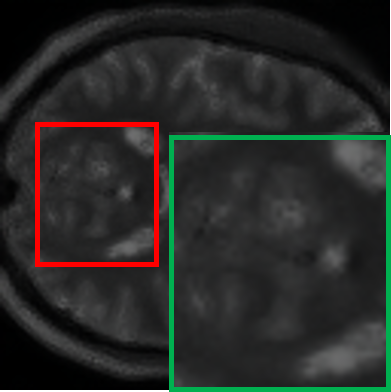}}\end{minipage}&\begin{minipage}[b]{0.4\columnwidth}\centering \raisebox{-.5\height}{\includegraphics[width=\linewidth]{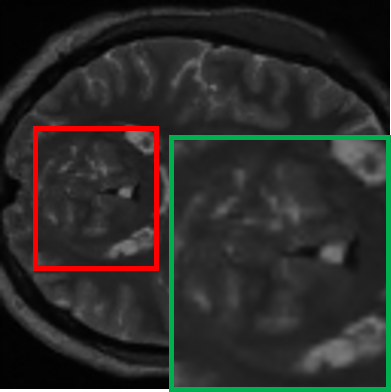}}\end{minipage}\\
\bottomrule
\end{tabular}
\end{threeparttable}
\end{table*}

\subsubsection{Quantitative metric-based analysis}
\begin{figure}[htbp]
\centering
\subfloat[]{
\label{PSNRR}
\centering
\includegraphics[width=0.23\textwidth]{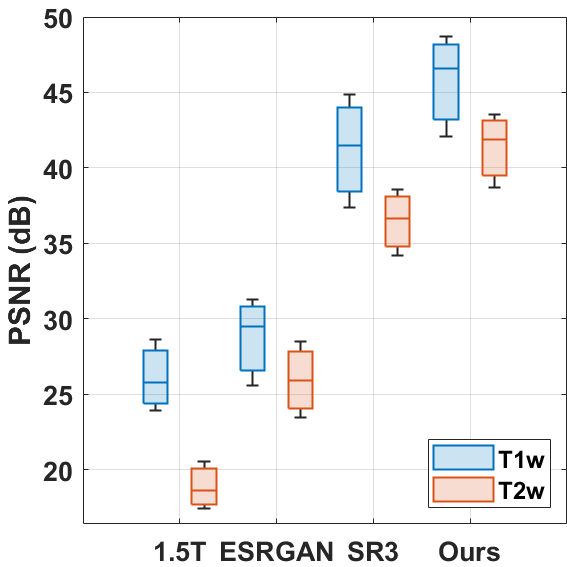}
}
\subfloat[]{
\label{SSIM}
\centering
\includegraphics[width=0.23\textwidth]{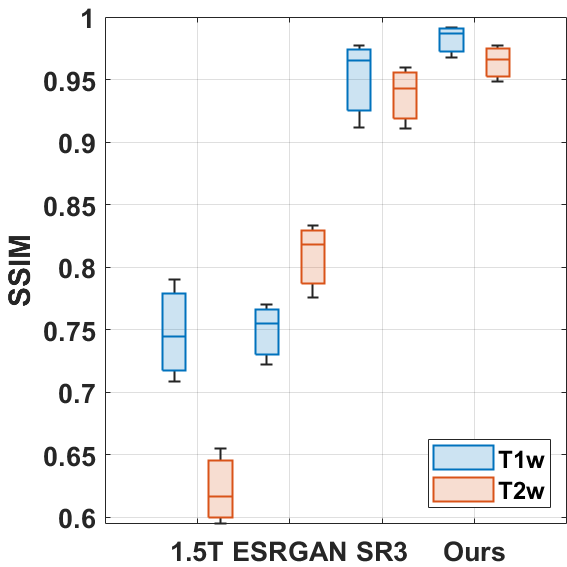}
}
\caption{The box plots visualize the different performance of the evaluated approaches using PSNR \ref{PSNRR} and SSIM \ref{SSIM}.}
\label{boxplots}
\end{figure}

Fig. \ref{boxplots} presents the statistical distributions of PSNR and SSIM metrics across various methods. The results are presented separately for T1w and T2w as they exhibit significant differences in image quality under 1.5T and 7T conditions. From the PSNR box plot (Fig. \ref{PSNRR}), it is evident that our proposed method achieves the highest median PSNR for both T1w and T2w modalities, exceeding 45 dB and 40 dB, respectively, indicating superior image quality reconstruction. In contrast, ESRGAN and SR3 exhibit lower medians, with SR3 outperforming ESRGAN but both trailing behind our method. The 1.5T baseline shows the lowest PSNR values, emphasizing the improvement achieved by super-resolution methods. Similarly, the SSIM box plot (Fig. \ref{SSIM}) reveals that our method attains the highest median SSIM for both modalities, approaching nearly 0.98, which reflects near-perfect structural similarity. SR3 also performs well, with slightly lower medians compared to our method, while ESRGAN and the 1.5T baseline display notably lower SSIM values.

Overall, these results collectively highlight the robust performance of our approach, and the efficacy of our approach in both preserving structural fidelity and enhancing image quality over competing methods.

\subsection{Selection of the step count for student model}
\label{step_count}
In the progressive distillation framework, the step count $N$ represents the total number of steps in the 1.5T-to-7T SR task. Adjusting $N$ directly affects the granularity of the student model's intermediate targets, influencing both the distillation process's effectiveness and computational complexity. Table \ref{vlisualization_200} highlights the impact of varying $N$ on the distillation framework, visualized in a 3D embedding space using t-SNE \cite{van2008visualizing}. The points include the real 1.5T data (a blue point), the middle-step reconstructed data (small grey points), and the 7T-like output data of the teacher model (a red point). As can be seen, as $N$ increases, the student model benefits from more intermediate subgoals, enabling finer-grained guidance for resolution refinement. Specifically, for $N < 20$, the limited intermediate steps force the student model to make abrupt, large transitions, leading to difficulty in denoising and refining structural details. Conversely, increasing $N$ to 50 smoothens the refinement trajectory, allowing for smaller incremental changes and better structural preservation. This trend is quantified by the decreasing PS percentage differences shown below each plot. As can be seen, for $N = 20$, the student model achieves around 90$\%$ of the teacher model's performance in PSNR. The computational and inference cost associated with higher $N$ becomes substantial, and the minor improvements in quality no longer justify the added overhead. It is noteworthy that the computational complexity of the distillation process is not solely linear with respect to $N$, it is also significantly influenced by the structural design of the student model. In theory, increasing $N$ could be balanced by employing a lighter student model, as the reduced per-step computational demands would offset the increase in the number of refinement steps. However, extensive experimentation in our study revealed that overly lightweight student models face substantial limitations in their denoising capabilities. Specifically, even with larger $N$ values, such models struggle to refine noised latent representations into high-quality outputs, which arises from their inability to capture complex structural details and high-frequency information. Considering these factors, $N = 20$ emerges as the practical choice for this study. This configuration strikes a balance between sufficient refinement stages and computational efficiency, ensuring that the student model has enough capacity to handle the intermediate targets while avoiding diminishing returns and excessive costs associated with larger $N$ values.

\begin{table*}[htbp]
\centering
\caption{Visualization of student model paths under different step counts $N$}
\label{vlisualization_200}
\setlength{\tabcolsep}{0.01mm}
\begin{threeparttable}
\begin{tabular}{ccccc}
\toprule
$N$=5 & $N$=10 & $N$=20& $N$=30 & $N$=50\\
\midrule
\begin{minipage}[b]{0.40\columnwidth}\centering \raisebox{-.5\height}{\includegraphics[width=\linewidth]{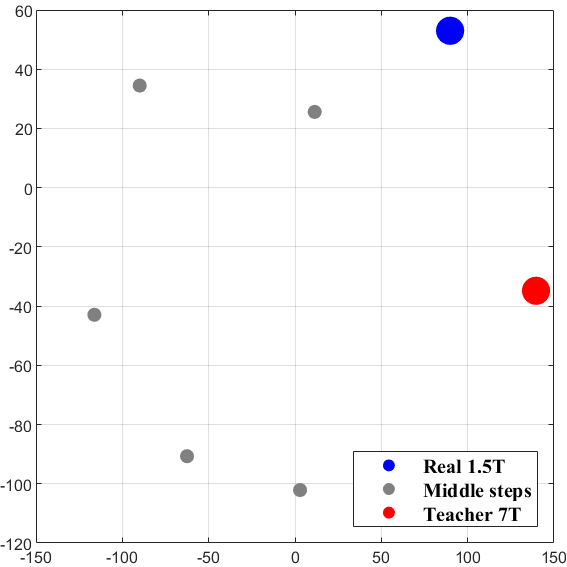}}\end{minipage}&\begin{minipage}[b]{0.40\columnwidth}\centering \raisebox{-.5\height}{\includegraphics[width=\linewidth]{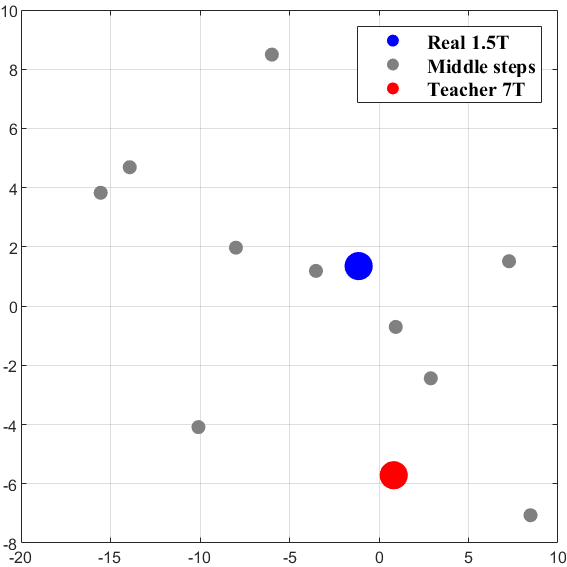}}\end{minipage}&\begin{minipage}[b]{0.40\columnwidth}\centering \raisebox{-.5\height}{\includegraphics[width=\linewidth]{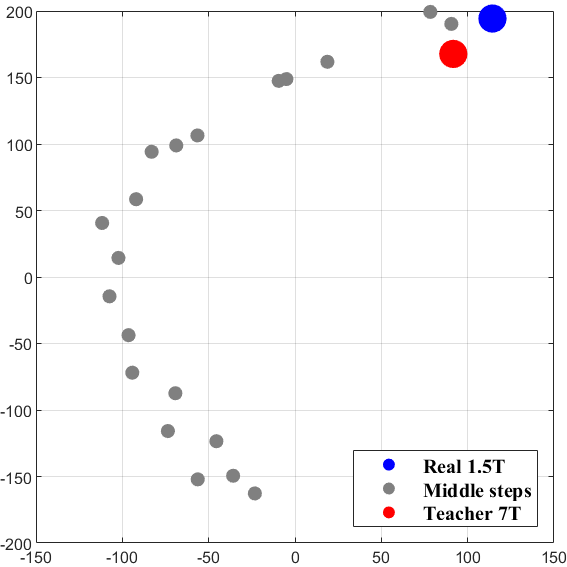}}\end{minipage}&\begin{minipage}[b]{0.40\columnwidth}\centering \raisebox{-.5\height}{\includegraphics[width=\linewidth]{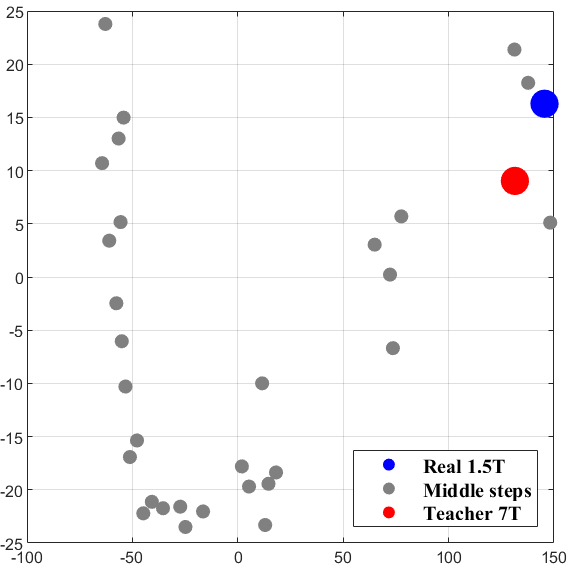}}\end{minipage}&\begin{minipage}[b]{0.40\columnwidth}\centering \raisebox{-.5\height}{\includegraphics[width=\linewidth]{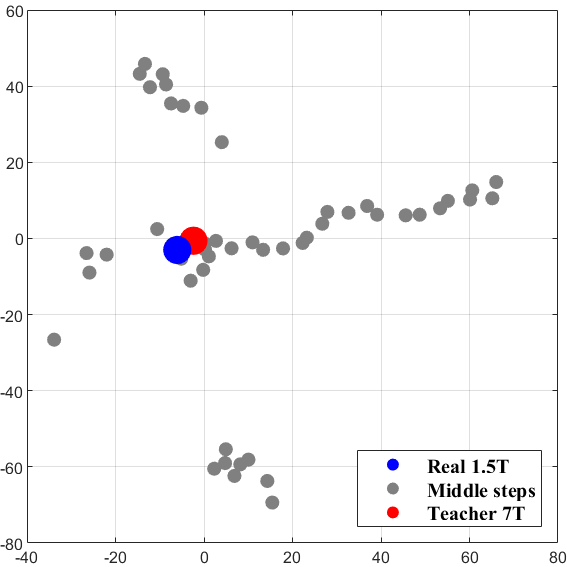}}\end{minipage}\\
\midrule
PSNR: 6.29 (86$\%\downarrow$) & PSNR: 18.31 (59$\%\downarrow$) & PSNR: 39.03 (13$\%\downarrow$) & PSNR: 39.88 (12$\%\downarrow$) & PSNR: 40.36 (11$\%\downarrow$)\\
\bottomrule
\end{tabular}
\begin{tablenotes}
\footnotesize
\item[$*$] According to the Fig. \ref{boxplots}, the average PSNR of 45.11 (dB) is as the teacher reference.
\end{tablenotes}
\end{threeparttable}
\end{table*}

As shown in Table \ref{Comparison_teacher_student}, the student model demonstrates a significant reduction in computational complexity and parameter count compared to the teacher model. Specifically, the student model achieves a 59$\%$ reduction in FLOPs and a 69$\%$ reduction in parameter number, underscoring its efficiency. Despite this substantial reduction, the student model retains competitive performance, with only a 13$\%$ decrease in PSNR (from 45.11dB to 39.03dB) and a 6$\%$ decrease in SSIM (from 0.96 to 0.90), which demonstrates the effectiveness of the student model in maintaining high-quality super-resolution results while significantly reducing resource requirements. Moreover, these results reflect the effectiveness of the progressive distillation framework, making it more suitable for deployment in resource-constrained environments.

\begin{table}[htbp]
\centering
\caption{Comparison of the teacher and student models}
\label{Comparison_teacher_student}
\setlength{\tabcolsep}{1.2mm}
\begin{tabular}{lcc|cc}
\toprule
Models & FLOPs ($G$) & Params ($M$)&  PSNR (dB) & SSIM\\
\midrule
Teacher & 92.59 &1156.37& 45.11&0.96\\
Student & 38.42 (59$\%\downarrow$)& 363.81 (69$\%\downarrow$) & 39.03 (13$\%\downarrow$) & 0.90 (6$\%\downarrow$)\\
\bottomrule
\end{tabular}
\end{table}

\subsection{Deployable flexibility analysis of the student model}
Beyond its lightweight architecture, the student model demonstrates a remarkable deployable ability to perform SR tasks across MRIs of varying resolutions without requiring additional training. This stems from its training process that involves exposing the student model to inputs and targets across a range of resolutions, equipping it with the ability to handle diverse resolution scales. For this, we explore two common SR scenarios: 1.5T to 3T and 3T to 7T MRI. We include the recently proposed Implicit Diffusion Model (IDM) \cite{gao2023implicit}, a novel method for continuous image SR, as a benchmark in our evaluation. In our study, the voxel sizes of the MRI scans are 2mm$\times$2mm$\times$2mm for 1.5T, 1.5mm$\times$1.5mm$\times$1.5mm for 3T, and 0.7mm$\times$0.7mm$\times$0.7mm for 7T. Since the focus is on 2D slice-based SR, we configure both the magnification factors for IDM and the step allocations for our proposed student model to match the required SR ratios. Specifically, for the 1.5T to 3T scenario, a magnification factor of 1.8 is employed for IDM, corresponding to an allocation of 4 steps out of the total 20 steps for the student model. Similarly, for the 3T to 7T scenario, a magnification factor of 4.6 is applied for IDM, corresponding to an allocation of the remaining 16 steps for the student model.

\begin{table*}[htbp]
\centering
\caption{Deployable flexibility analysis of the student model}
\label{deployable}
\setlength{\tabcolsep}{0.4mm}
\begin{threeparttable}
\begin{tabular}{ccccc}
\toprule
\multicolumn{5}{c}{\bf {Scenario 1: from 1.5T to 3T-like}}\\
Input (1.5T) & Ground Truth (3T) &Our teacher& IDM (1.8$\times$)& Our student$^1$\\
\midrule
\begin{minipage}[b]{0.4\columnwidth}\centering \raisebox{-.5\height}{\includegraphics[width=\linewidth]{T1w_sagi_1_5T.png}}\end{minipage}&\begin{minipage}[b]{0.4\columnwidth}\centering \raisebox{-.5\height}{\includegraphics[width=\linewidth]{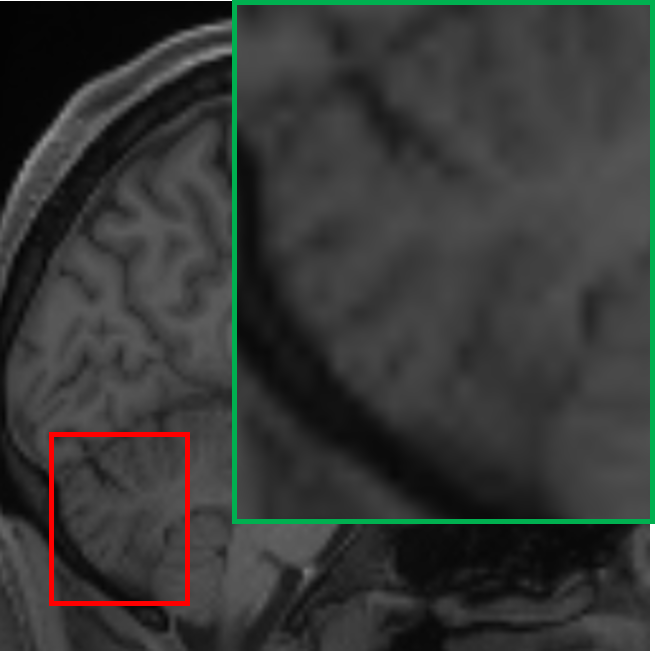}}\end{minipage}&
\diagbox[width=2cm, height=2cm, dir=SW]{}{} &\begin{minipage}[b]{0.4\columnwidth}\centering \raisebox{-.5\height}{\includegraphics[width=\linewidth]{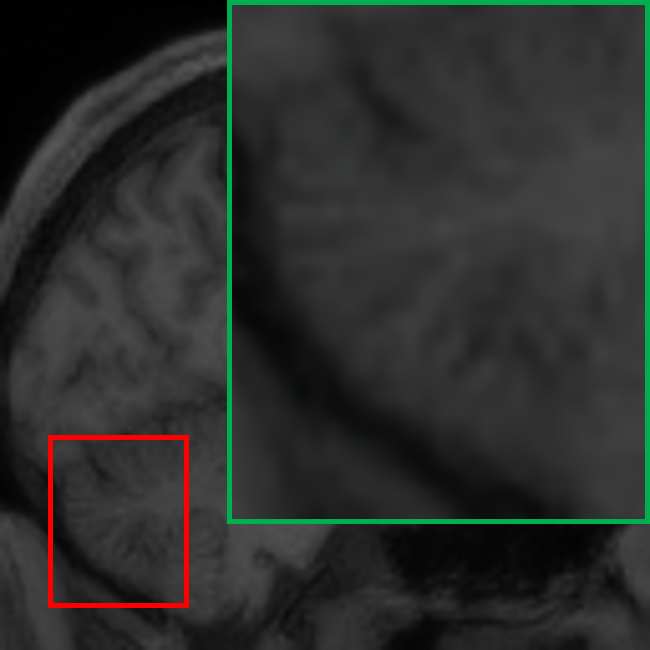}}\end{minipage}&\begin{minipage}[b]{0.4\columnwidth}\centering \raisebox{-.5\height}{\includegraphics[width=\linewidth]{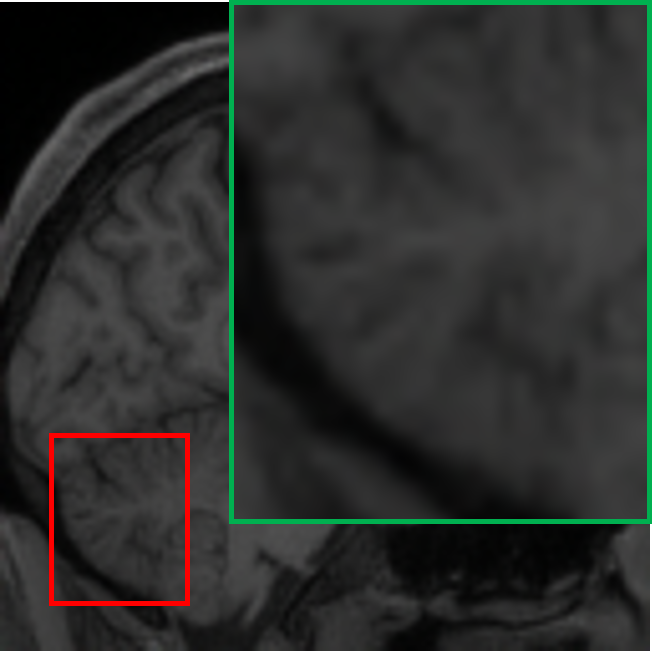}}\end{minipage}\\
\midrule
\multicolumn{5}{c}{\bf {Scenario 2: from 3T to 7T-like}}\\
Input (3T) & Ground Truth (7T) &Our teacher & IDM (4.6$\times$) & Our student$^2$\\
\midrule
\begin{minipage}[b]{0.4\columnwidth}\centering \raisebox{-.5\height}{\includegraphics[width=\linewidth]{T1w_sagi_3T.png}}\end{minipage}&\begin{minipage}[b]{0.4\columnwidth}\centering \raisebox{-.5\height}{\includegraphics[width=\linewidth]{T1w_sagi_7T.png}}\end{minipage}&\begin{minipage}[b]{0.4\columnwidth}\centering \raisebox{-.5\height}{\includegraphics[width=\linewidth]{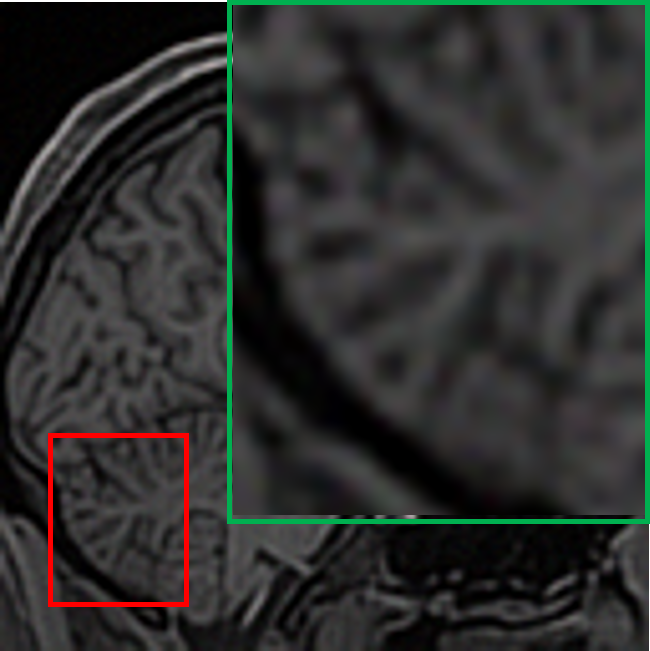}}\end{minipage}&\begin{minipage}[b]{0.4\columnwidth}\centering \raisebox{-.5\height}{\includegraphics[width=\linewidth]{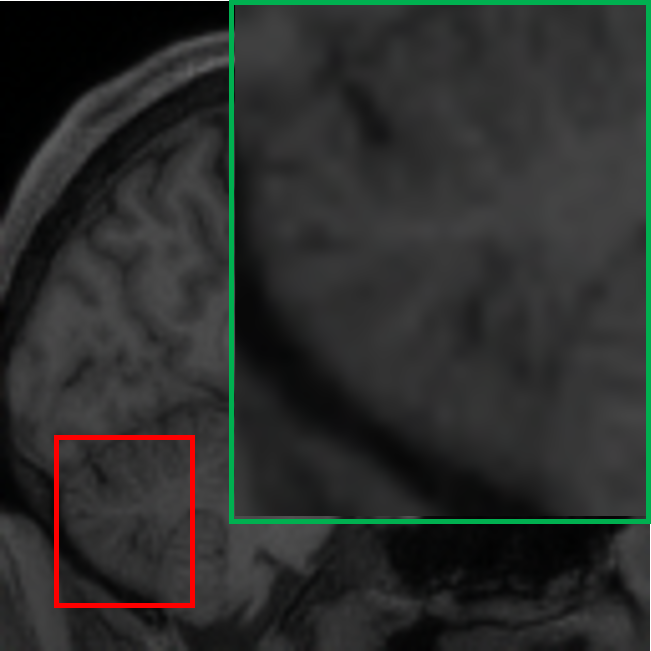}}\end{minipage}&\begin{minipage}[b]{0.4\columnwidth}\centering \raisebox{-.5\height}{\includegraphics[width=\linewidth]{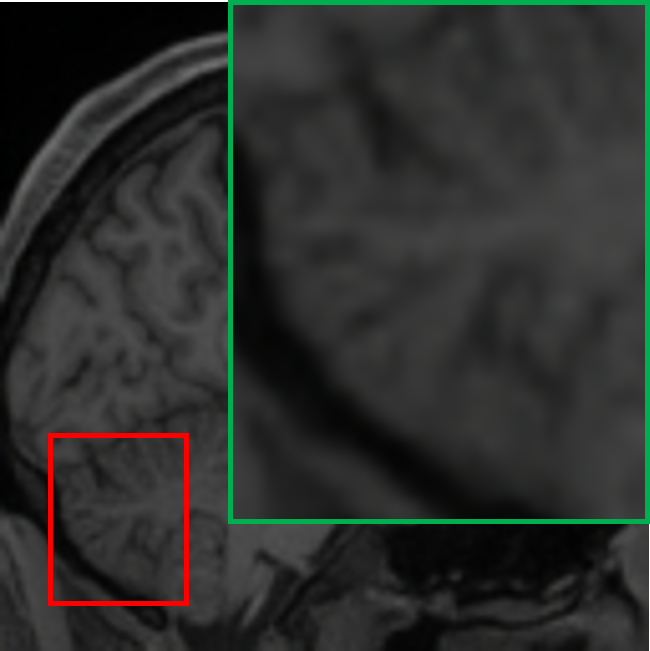}}\end{minipage}\\
\bottomrule
\end{tabular}
\begin{tablenotes}
\footnotesize
\item[$^1$] Student model step allocations $N$:1$\sim$4 $|$ $N$=20
\item[$^2$] Student model step allocations $N$:5$\sim$20 $|$ $N$=20
\end{tablenotes}
\end{threeparttable}
\end{table*}

As illustrated in Figure \ref{deployable}, for the 1.5T-to-3T task, the teacher model is not applicable as it is designed to directly perform SR from 1.5T to 7T. The output of IDM in this scenario shows significant blurring, particularly in the cerebellum region, with minimal perceptible improvement over the input, indicating a limitation in capturing the finer structures. In contrast, our student model achieves results that closely resemble the 3T ground truth, effectively recovering structural details and textures. For the 3T-to-7T task, the teacher model, when directly applied to 3T data, generates overly sharpened outputs, reflecting its sensitivity to input-specific constraints and its lack of adaptability to intermediate resolutions. However, the student model consistently outperforms IDM in this scenario as well, delivering results with better alignment to the 7T ground truth and more balanced enhancement of fine details. These observations highlight the flexibility and robustness of the student model across varying resolution scales, emphasizing its superiority in deployable SR scenarios.

\subsection{Clinical evaluation}
To ensure the clinical validity of our proposed SR framework, ethical approval was obtained from the Institutional Review Board (IRB) at Massachusetts General Hospital (MGH) under protocol number 2024P003489. All datasets used in this study were fully de-identified to adhere to strict ethical guidelines and privacy regulations.

\begin{table*}[htbp]
\centering
\caption{Clinical evaluation for seizure case}
\label{clinical_1}
\setlength{\tabcolsep}{0.4mm}
\begin{tabular}{ccc}
\toprule
 Real 1.5T MRI &  Real 3T MRI &  3T-like SR MRI (1.5T to 3T)\\
\begin{minipage}[b]{0.65\columnwidth}\centering \raisebox{-.5\height}{\includegraphics[width=\linewidth]{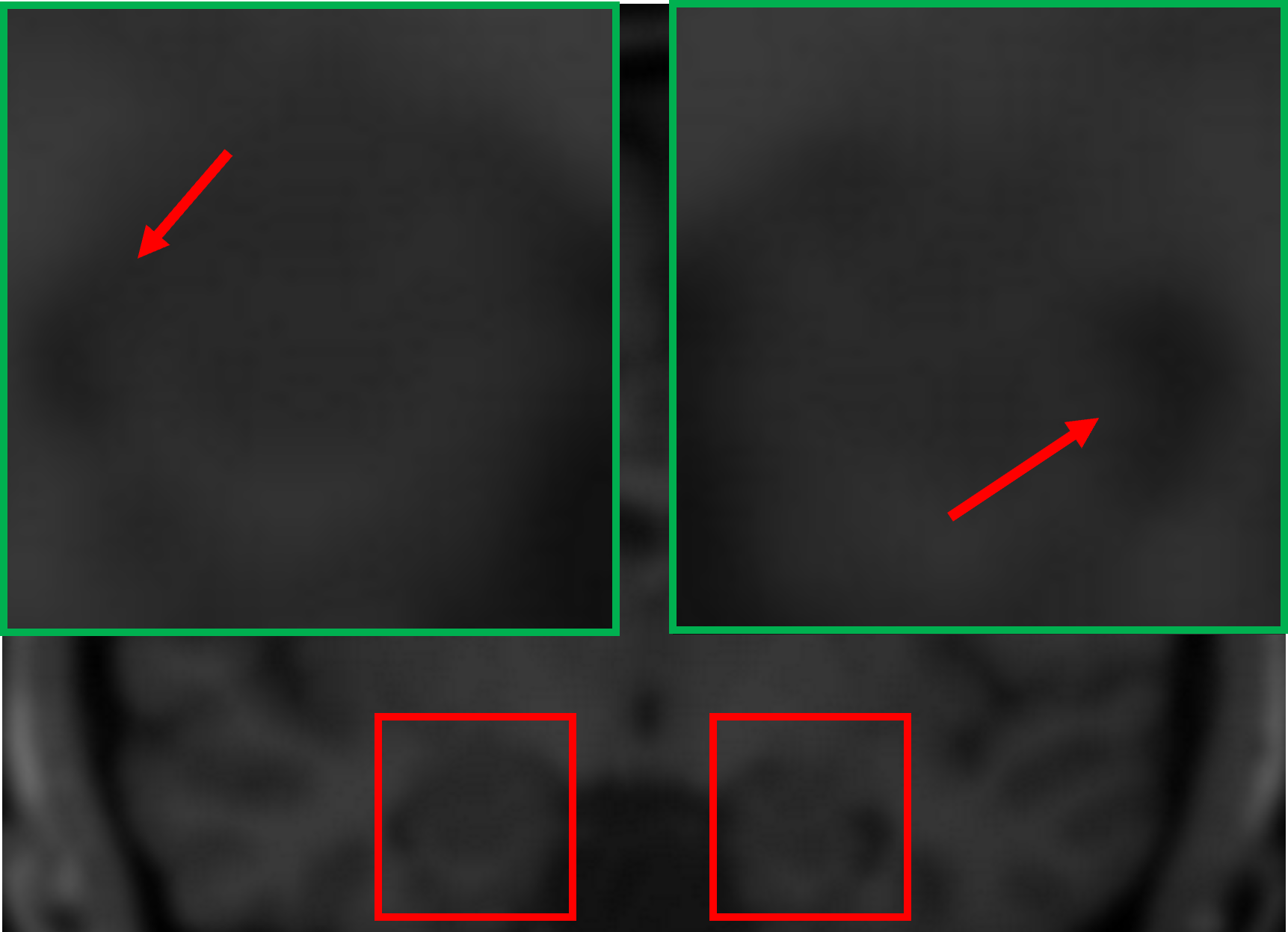}}\end{minipage}&\begin{minipage}[b]{0.65\columnwidth}\centering \raisebox{-.5\height}{\includegraphics[width=\linewidth]{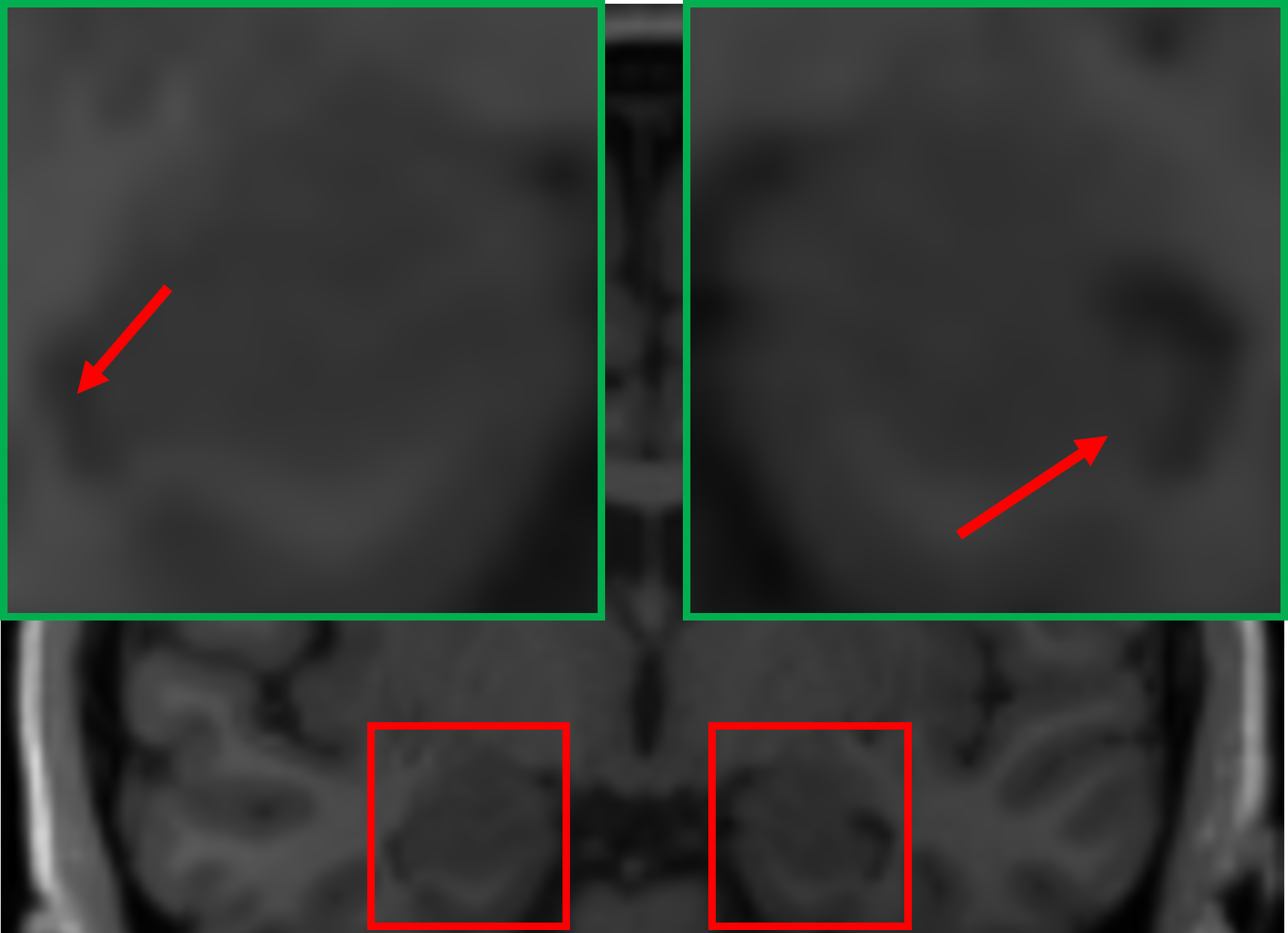}}\end{minipage}&\begin{minipage}[b]{0.65\columnwidth}\centering \raisebox{-.5\height}{\includegraphics[width=\linewidth]{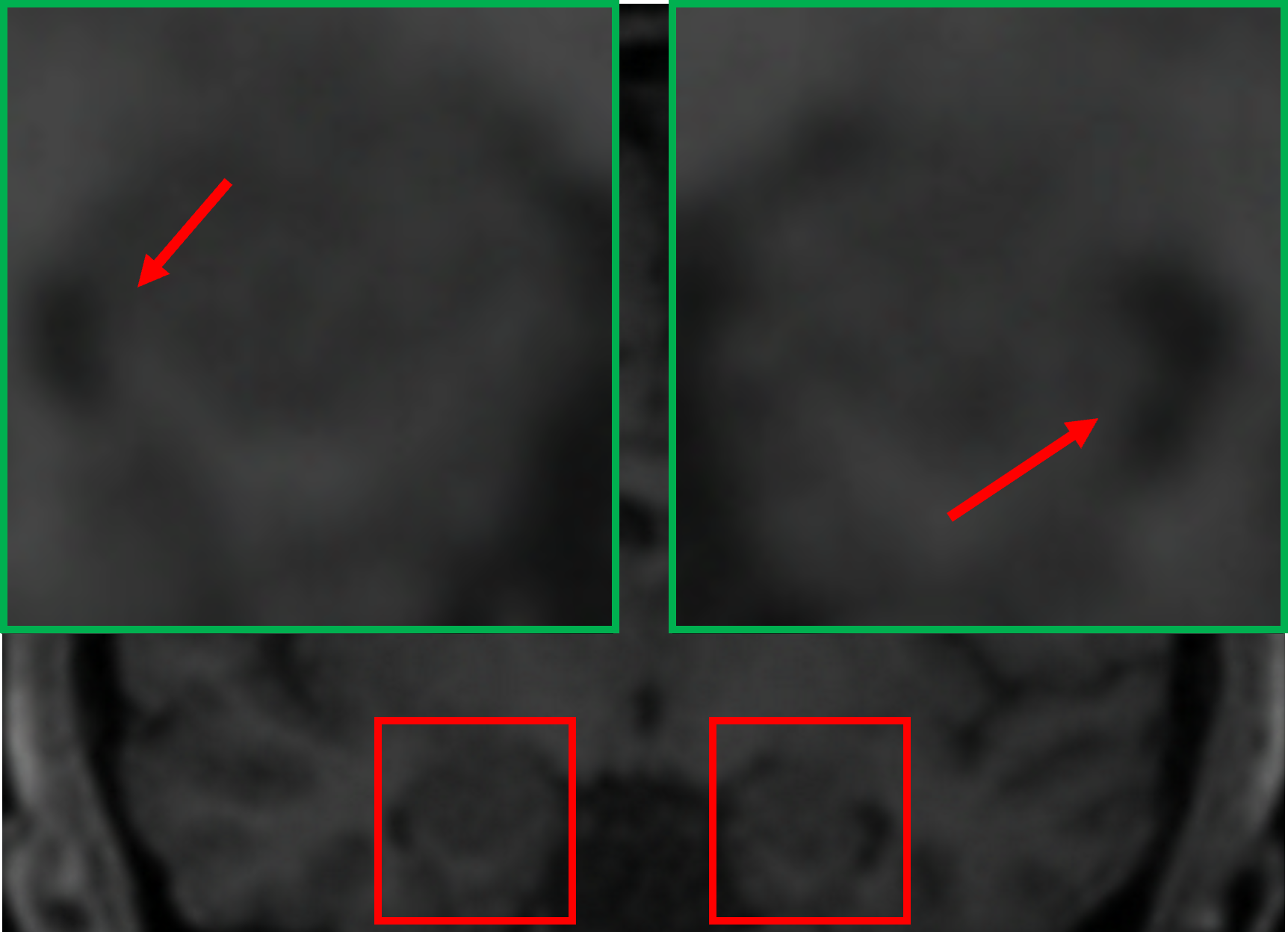}}\end{minipage}\\
\bottomrule
\end{tabular}
\end{table*}

In one case, for a patient with partial seizures and secondary generalization, an initial diagnosis using 1.5T MRI revealed a mild volume loss in the left hippocampus, further imaging with 3T MRI was recommended to confirm the diagnosis. As shown in Table \ref{clinical_1}, the green boxes highlight the hippocampus region, while the red arrows point to areas indicative of water content within the hippocampus. In the 1.5T MRI, the patient’s hippocampal asymmetry can be appreciated with large sulci with water content on the left side (right sub-image), suggesting mesial temporal lobe atrophy. This observation was confirmed more clearly in the subsequent 3T MRI, which provided a sharper depiction of the sulcal enlargement in the same region, reinforcing the diagnosis of mesial temporal disease. Notably, by applying our student model to perform SR on the 1.5T MRI, the resulting 3T-like MRI closely resembled the actual 3T MRI. The obtained image not only improved the clarity of the hippocampal structure but also effectively captured the increased water content in the left hippocampus, closely matching the pathology observed in the 3T MRI, which underscores the clinical potential of our super-resolution framework in enhancing diagnostic capabilities. By applying our student model to super-resolve the 1.5T MRI to a 3T-like resolution, the resulting image closely resembled the actual 3T MRI. The 1.5T-based SR image not only improved the clarity of the hippocampal structure but also effectively captured the increased water content in the left hippocampus, closely matching the pathology observed in the 3T MRI.
\begin{table*}[htbp]
\centering
\caption{Clinical evaluation for MS case}
\label{clinical_2}
\setlength{\tabcolsep}{0.4mm}
\begin{tabular}{ccc}
\toprule
 Real 3T MRI &  Real 7T MRI &  7T-like SR MRI (3T to 7T)\\
\begin{minipage}[b]{0.65\columnwidth}\centering \raisebox{-.5\height}{\includegraphics[width=\linewidth]{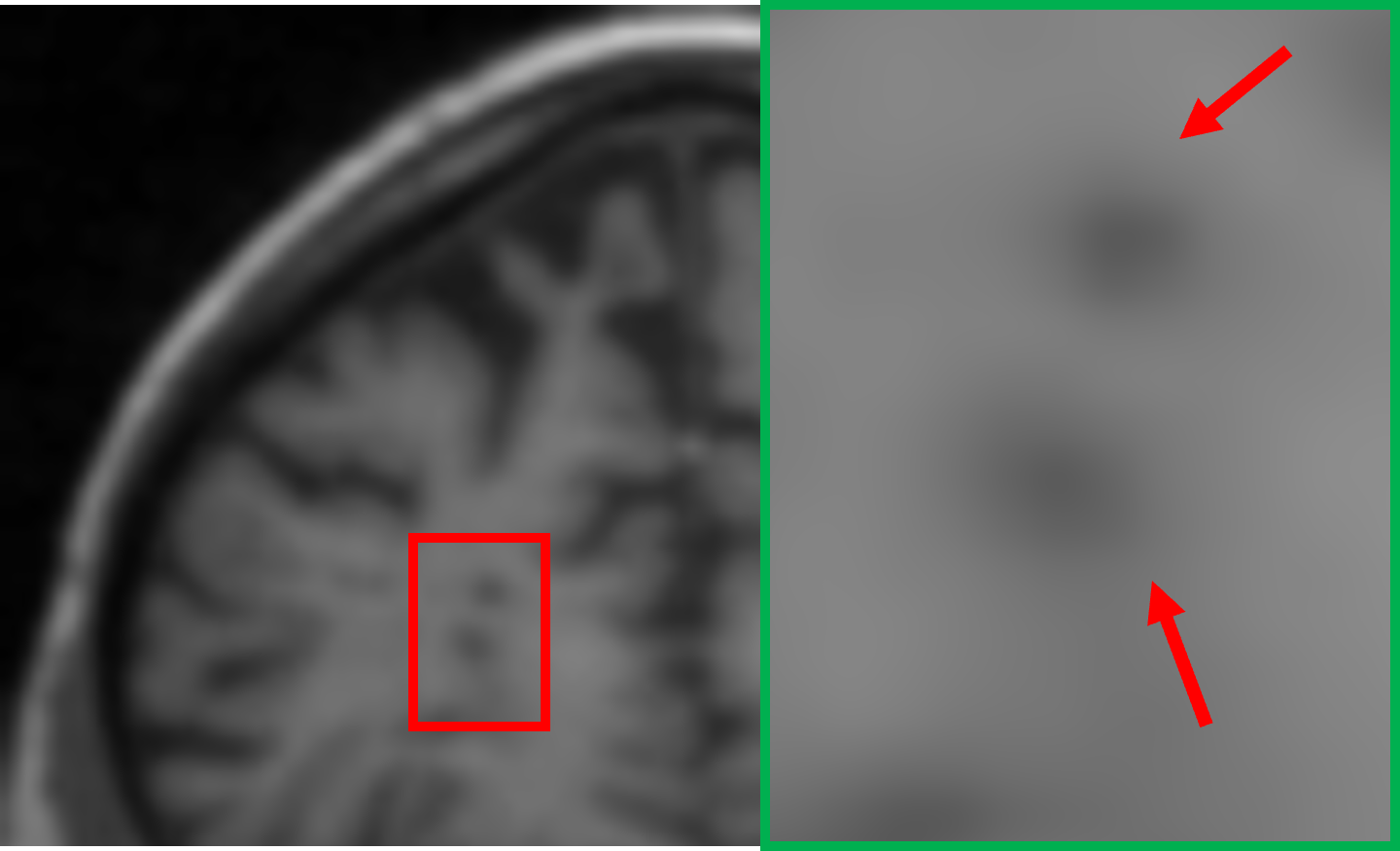}}\end{minipage}&\begin{minipage}[b]{0.65\columnwidth}\centering \raisebox{-.5\height}{\includegraphics[width=\linewidth]{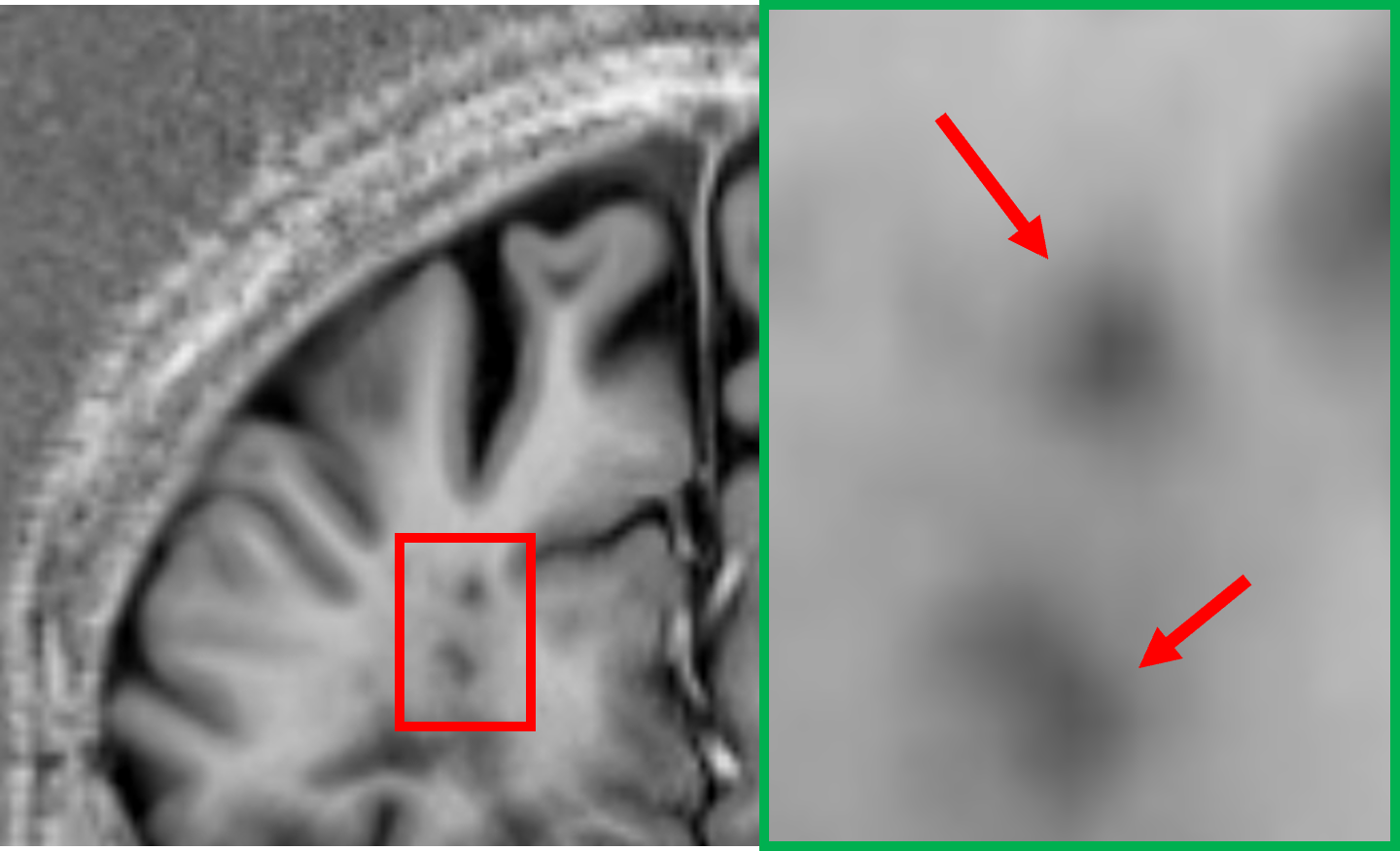}}\end{minipage}&\begin{minipage}[b]{0.65\columnwidth}\centering \raisebox{-.5\height}{\includegraphics[width=\linewidth]{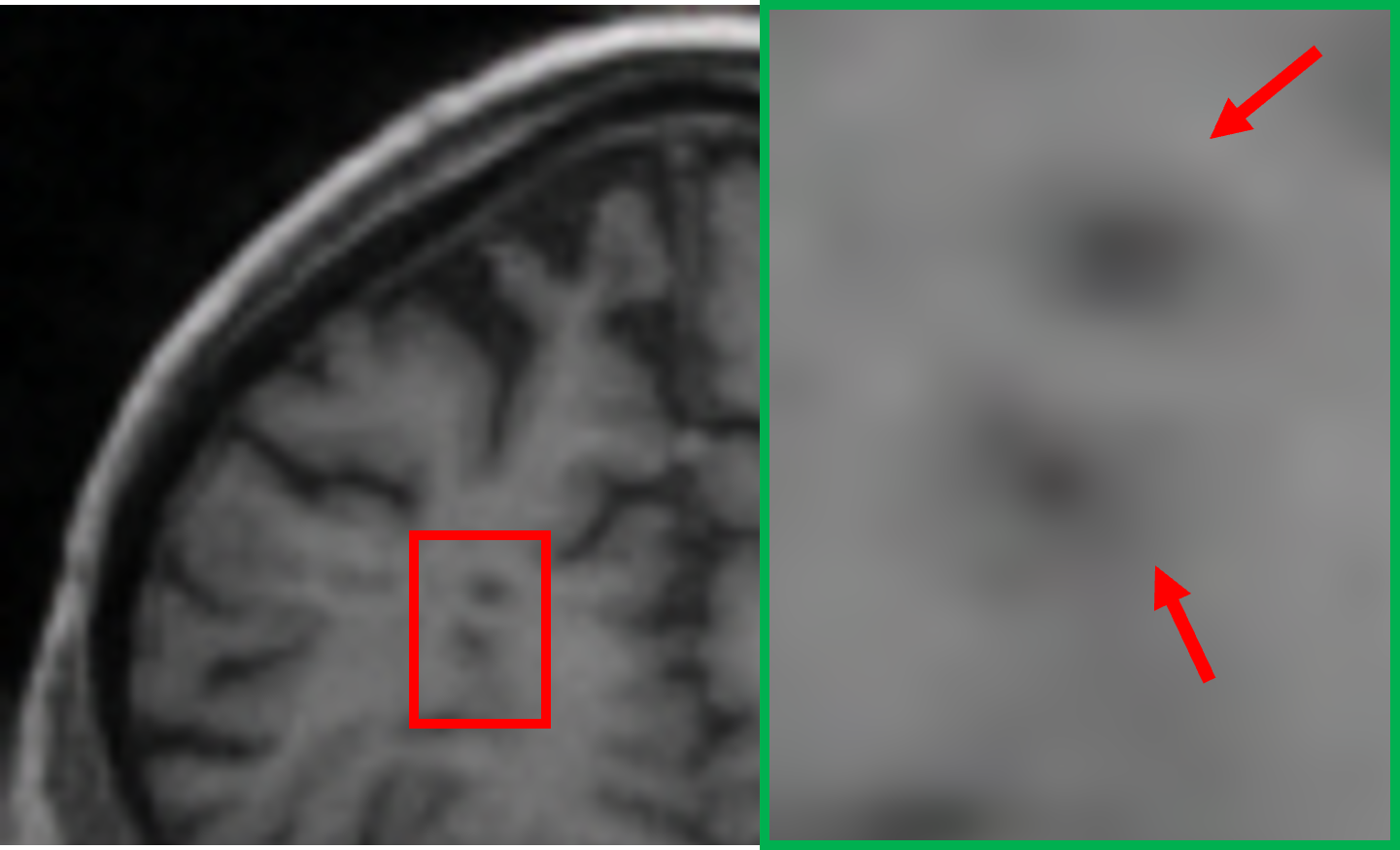}}\end{minipage}\\
\bottomrule
\end{tabular}
\end{table*}

In another case involving a patient with multiple sclerosis (MS), characterized by periventricular and subcortical white matter abnormalities. MS lesions, often appearing as small “dark holes” in the brain on T1w MRI, are notoriously difficult to identify on lower-field MRIs. As shown in Table \ref{clinical_2}, the real 7T MRI provided a significant improvement in resolution compared to the 3T MRI, enabling clearer visualization of MS-related changes in the periventricular regions, which highlights the importance of 7T MRI in detecting subtle lesions. By applying our student model to super-resolve the 3T MRI into a 7T-like quality, the resulting SR image closely matched the real 7T MRI, markedly improving the depiction of the periventricular white matter, making the small “dark holes” indicative of MS lesions much more apparent.

The above clinical cases underscore the clinical value and potential of our proposed SR framework. In the seizure case, our model attempted to bridge the gap between 1.5T and 3T imaging, enabling the detection of subtle hippocampal atrophy with improved clarity. Similarly, in the MS case, the model showed promise in enhancing lesion conspicuity, effectively highlighting subtle abnormalities. These findings show that by super-resolving MRIs to higher-quality ones, our approach demonstrates the potential to enhance diagnostic capabilities in resource-limited settings.

\subsection{Discussion}
This study introduces a novel framework for super-resolution (SR) MRI, leveraging conditional diffusion models guided by bias field correction and gradient nonlinearity correction to generate 7T-like MRI volumes from 1.5T inputs. The proposed model, through comprehensive comparisons with state-of-the-art (SOTA) models, demonstrates superior performance. However, we did not stop there. A progressive distillation strategy is introduced, enabling a lightweight student model to approximate the high-quality outputs of a larger teacher model while significantly reducing computational requirements. This innovation not only enhances the practical deployability of SR models in resource-constrained environments but also ensures adaptability across varying MRI resolutions, making the framework versatile and scalable. Clinical evaluations further validate the framework's utility, demonstrating that the SR images are not only visually accurate but also diagnostically meaningful. These results underscore the potential of SR models to bridge the resolution gap, providing enhanced diagnostic capabilities in scenarios where high-field MRIs may not be available. Several critical aspects of the study merit further discussion. 

\subsubsection{How it works?}
In this study, we shift our focus from the differences in imaging field strengths and scanning configurations between 1.5T and 7T MR scans. Instead, we concentrate on the intricate mapping relationship between these two resolution levels. Similar to how an experienced radiologist can mentally infer structural and contrast details from lower-resolution scans, our artificial intelligence-based (AI-based) model is trained on high-quality paired data (i.e., 1.5T and 7T MR scans) to accurately learn these mapping relationships, achieving such inference.

\subsubsection{Choice of teacher model outputs as subgoal references}
A notable design choice is the use of 7T-like outputs from the teacher model, rather than ground truth 7T data, as the reference for progressive distillation, aligning the student model's learning trajectory with the teacher's distilled knowledge, enabling the global guidance from the feature maps of teacher model and consistency within the progressive distillation framework.

\subsubsection{Strengths and limitations}
\label{plusandlimi}
This study has several notable strengths. Firstly, it introduces an innovative approach for generating HR MRI volumes from LR MRI inputs using advanced diffusion-based models and explores the potential to address the resolution gap between different field strengths. Our experiments demonstrate that incorporating supplementary information, such as bias field correction and gradient nonlinearity correction, significantly improves the quality of the generated super-resolved MRIs, which underscores the potential of integrating diverse correction mechanisms to produce more clinically valuable outputs. Secondly, to further enhance the deployability of the super-resolution model, we developed a progressive distillation strategy that enables the student model to gradually refine its outputs toward 7T-like quality through stepwise learning. The student model, trained under this framework, achieves high-quality outputs with a significantly lighter architecture, making it practical for real-world deployment. Finally, the study emphasizes pathology-specific assessments by expert radiologists, which ensures that the generated super-resolved MRIs are not only visually faithful to ground truth but also diagnostically meaningful. Despite these strengths, the study has several limitations. Firstly, the performance of the model is heavily dependent on the availability of large, high-quality paired data. The scarcity of datasets, especially for rare conditions or specialized applications, may restrict the generalizability of the approach. Secondly, although the progressive distillation strategy results in a significantly lighter student model, the student model still requires approximately 15GB of graphic memory, deploying the framework on common computers thus remains challenging. Finally, the guidance of bias field correction and gradient nonlinearity correction, while beneficial for generating high-quality outputs, introduces an additional dependency on pre-processing steps, which could pose challenges in workflows where these corrections are unavailable or impractical to implement consistently.

\subsubsection{Future work}
Future work for this study will focus on expanding the generalizability and robustness of the proposed SR framework. One key direction is the exploration of multi-modal data integration, combining MRI with complementary imaging modalities such as CT or ultrasound, to improve the reconstruction process and accuracy. Additionally, expanding the framework to accommodate other imaging modalities could further broaden its applicability and impact across diverse clinical and research domains.

\section{Statements}
This manuscript was prepared using data from the HCP project. The views expressed in it are those of the authors and do not necessarily reflect the opinions of the HCP investigators, the National Institutes of Health (NIH), or the private funding partners.

\section{Acknowledgements}
The authors gratefully acknowledge the support of the Ralph Schlaeger Research Fellowship under award number 246448 from Massachusetts General Hospital (MGH), Harvard Medical School (HMS) and the French National Agency of Research (ANR) under project number ANR-20-CE45-0013-01.

\bibliographystyle{IEEEtran}
\bibliography{references}

\end{document}